\documentclass{gandb}
\usepackage{amsmath}
\usepackage{graphicx}
\usepackage{epsfig}
\usepackage{float}

\def\eqref#1{(\ref{#1})}
\let\<=\langle
\let\>=\rangle

\begin{document}
\title{ENVELOPE OSCILLATIONS OF INTENSE BEAMS}
{THEORETICAL STUDIES OF\\
ENVELOPE OSCILLATIONS\\
AND INSTABILITIES OF MISMATCHED\\
INTENSE CHARGED-PARTICLE BEAMS\\
IN PERIODIC FOCUSING CHANNELS}
\author{J.~STRUCKMEIER}
\authorhead{J.~STRUCKMEIER AND M.~REISER}
\address{\it Gesellschaft f\"ur Schwerionenforschung (GSI),
Postfach~110541,\\ 6100~Darmstadt, W.~Germany\\~\\
{\rm and}}
\author{M.~REISER}
\address{Electrical Engineering Department and Department of Physics and Astronomy, University of Maryland, College Park, Maryland 20742, U.S.A.}
\received{(Received May 18, 1983; in final form June 6, 1983; revised 16 October 2026)}
\abstract{The behavior of mismatched intense charged-particle beams in periodic transport channels of the solenoid and quadrupole type is studied theoretically.
The envelope-oscillation frequencies of the mismatched beam are obtained by the smooth-approximation method and by numerical evaluation of the linearly perturbed K-V envelope equations.
Phase shifts of the envelope oscillations and growth factors per focusing period in the case of instability are calculated for a solenoid and a magnetic quadrupole (FODO) channel using the parameters of the Maryland and GSI beam transport experiments.
For comparison and the purpose of illustration, the K-V equations are integrated numerically, and envelope curves as well as single-particle trajectories for mismatched beams are shown in graphical form.
In addition, computer simulation studies with the PARMILA code were performed, and results are presented both for K-V and a Gaussian distribution in transverse phase space.\\~\\
Revised version of: Particle~Accelerators~{\bf 14}, 227--260 (1984)}
\maketitle
\section{INTRODUCTION}
The transport of intense charged-particle beams through periodic focusing structures is of considerable interest for a variety of applications.
In particular, the possibility of using high-power heavy-ion beams for inertial fusion has triggered new investigations of this problem.
Theoretical studies of space-charge dominated beam transport usually involve the use of a self-consistent Kapchinskij-Vladimirskij (K-V) distribution,\cite{kapvla} for which the forces acting on the particles are linear functions of the displacement from the beam axis.
For a given focusing channel, the beam physics is largely determined by the phase advance (or ``betatron tune'') of single-particle oscillations in one channel period without space charge, $\sigma_0$ (which depends on the strength of the applied focusing forces) and the corresponding ``depressed'' phase advance $\sigma$ in the presence of space-charge forces.
The parameter dependence and scaling laws in the case of an ideal, matched beam have been studied by numerical integration (using scaled variables)\cite{lambertson} and by analytical solution (smooth approximation)\cite{reiser78} of the K-V envelope equations --- both for quadrupole and solenoid systems.
Instabilities have been investigated with the Vlasov equation for a K-V distribution and by numerical simulation.\cite{hofmann}
The main conclusion from these stability studies is that the phase advance $\sigma_0$ should not exceed $90^\circ$ (preferably $60^\circ$).
For $\sigma_0 > 90^\circ$, envelope (second-order) modes can become unstable in specific intensity intervals, even at relatively low intensities.
This contrasts with the situation where space-charge forces are negligible, in which case $\sigma_0$ can have values up to $180^\circ$.
For $\sigma_0\le 90^\circ$, modes of order $3$ and higher can impose additional restrictions on the phase advance $\sigma$ with space charge and hence on the transportable beam current.
However, there is a question whether real beams are affected by these modes, and several experiments are in progress to study this problem.\cite{reiser83,klabunde,chupp}

The amount of beam current that can be transported through a periodic focusing channel is a maximum when the beam is perfectly matched to the acceptance of the channel, provided that the betatron tunes $\sigma_0$ and $\sigma$ are chosen to avoid instabilities.
For a K-V beam, perfect matching implies that the envelope radii and the corresponding rms widths repeat after one focusing period.
Their period-averaged values remain constant, but the rms widths themselves generally have the periodic envelope ripple.
With a real beam, one usually seeks an approximate rms match.
Matching the initial rms parameters of a general non-K-V distribution to those of a matched K-V beam does not, by itself, make the full distribution a self-consistent periodic solution.
Furthermore, real beams often have a finite pulse length, and the intensity may vary within the pulse.
In this case, an input match for the central part of the pulse does not generally match the tails at the front and rear ends, where the intensity differs.
For this reason, an understanding of the behavior of mismatched beams is of great practical importance.
In our paper we present the results of a theoretical study of mismatched high-intensity beams in periodic solenoid and quadrupole channels.
The ongoing experiments at the University of Maryland and GSI Darmstadt provide the main focus for our work.
However, the framework and parameter range were chosen broadly enough so that the results are of general usefulness and scalable to other beam-transport systems.
We begin our analysis in the next section by linearizing the K-V envelope equations for a mismatched beam within the smooth approximation, which replaces the periodic lattice by equivalent continuous focusing.
By considering small perturbations of the known matched-beam solution and linearization of the equations, the two fundamental frequencies (or phase angles) for the oscillations of the mean radius are found.

In Section~\ref{sec3}, the periodic variation of the focusing forces is taken into account, and the eigenvalues (amplitude and phase angles) of the solutions to the linearized, perturbed K-V envelope equations are determined by numerical integration without smoothing the periodic focusing coefficients.
This part provides a direct link to the studies by Hofmann et~al.\ in Ref.~4, and our results confirm the existence of envelope instabilities in the region $\sigma_0 > 90^\circ$.
Below $\sigma_0 = 90^\circ$, no instability of the envelope modes considered here is found for the lattices studied, and the periodic-coefficient calculation and the smooth-approximation technique yield nearly identical envelope-oscillation frequencies.

In Section~\ref{sec4} we present computer results for the beam envelopes and single-particle trajectories under mismatched conditions for several cases of solenoid and quadrupole beam transport.
These results are obtained by numerical integration of the full nonlinear K-V equations.

Finally in Section~\ref{sec5}, the results of computer simulation studies for both \mbox{K-V} and Gaussian distributions are shown for a few typical situations relevant to the Maryland and GSI experiments.
The more realistic Gaussian distributions constitute a bridge between the ideal K-V model of the theory and the laboratory beam.
A comparison between the two distributions thus provides valuable help in obtaining an understanding of experimental data.

\section{SMOOTH APPROXIMATION THEORY OF BEAM-ENVELOPE OSCILLATIONS DUE TO MISMATCH\label{sec2}}

\begin{figure}[htb]
\epsfig{file=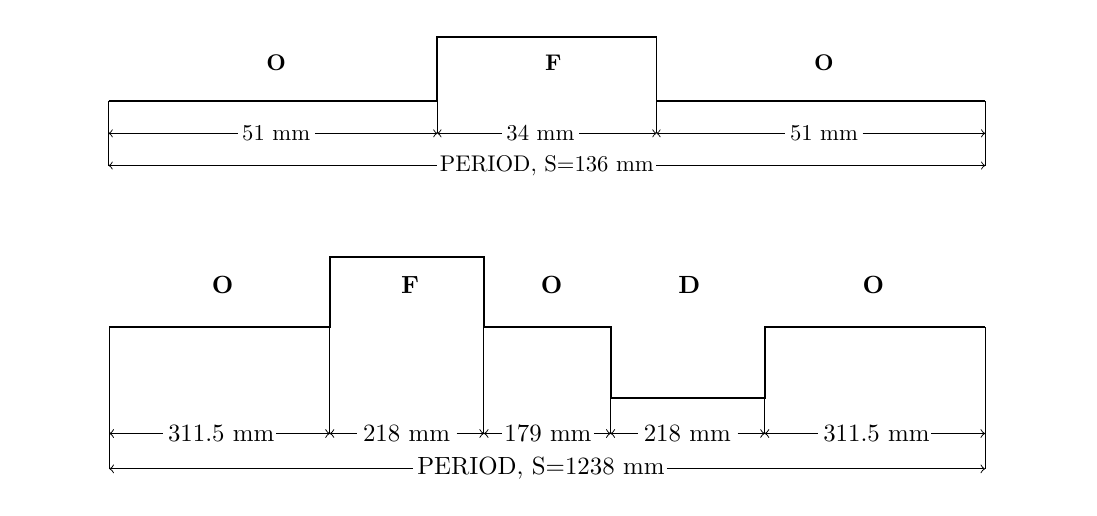,width=125mm,height=65mm}
\caption{Focusing functions $\kappa^2(s)$ for the Maryland solenoid channel (top) and for the GSI quadrupole channel (bottom).}
\label{fig1}
\end{figure}

We start our analysis with the well-known K-V differential equations for the beam envelopes $X(s)$ and $Y(s)$ in the two transverse directions as functions of distance~$s$ along the axis of the periodic focusing channel
\begin{subequations}\label{eq1}
\begin{eqnarray}
X^{\prime\prime}+\kappa_x{}^2(s)\cdot X-2K/(X+Y)-\epsilon^2/X^3&=&0,\label{eq1a}\\
Y^{\prime\prime}+\kappa_y{}^2(s)\cdot Y-2K/(X+Y)-\epsilon^2/Y^3&=&0.\label{eq1b}
\end{eqnarray}
\end{subequations}
Here $\kappa^2$ represents the signed applied periodic focusing coefficient.
For a rectangular (``hard-edge'') variation, as shown in Fig.~\ref{fig1}, it is given by
\begin{eqnarray*}
\kappa_x{}^2=\kappa_y{}^2=\kappa^2&=&{[qB/(2m_0c\beta\gamma)]}^2\qquad\mathrm{for\, solenoids,\, and}\\
\kappa_{x(y)}^2&=&\pm qB/(m_0c\beta\gamma a)\qquad\,\mathrm{for~magnetic~quadrupoles}
\end{eqnarray*}
of aperture radius $a$.
$K$ is the generalized perveance, defined as
\begin{displaymath}
K=(I/I_0)\cdot(2/\beta^3\gamma^3),\quad I_0=4\pi\epsilon_0m_0c^3/q,
\end{displaymath}
where $I_0\simeq 1.7\times 10^4$ Amp for electrons and $I_0\simeq 3.1\times 10^7~(A/Z)$ Amp for ions of mass number $A$ and charge $q = Ze$.
$\epsilon$ is the unnormalized K-V edge emittance ($\pi\epsilon =$ area of the ellipse in $(x,x')$-phase space), considered to be the same in both transverse directions for our studies.
For the uncoupled K-V distribution, the edge radii and emittance are related to the rms quantities by
\begin{displaymath}
 X=2x_{\mathrm{rms}},\qquad Y=2y_{\mathrm{rms}},\qquad
 \epsilon=4\epsilon_{x,\mathrm{rms}}=4\epsilon_{y,\mathrm{rms}},
\end{displaymath}
where $\epsilon_{x,\mathrm{rms}}^2=\langle x^2\rangle\langle x'^2\rangle-\langle xx'\rangle^2$ and the averages are over centered particle coordinates.
For solenoid focusing, Eqs.~(\ref{eq1}) are written in the rotating Larmor frame for an uncoupled K-V distribution with zero mean canonical angular momentum.
The emittances in these equations are the corresponding Larmor-frame invariant emittances, not in general the projected laboratory-frame emittances.

If we introduce the mean values $\bar X(s)$ and $\bar Y(s)$ for the envelope radius, which vary slowly with $s$ and the short-wavelength ``ripple'' functions $\delta_x(s)$ and $\delta_y(s)$, which are small compared to unity and have the periodicity of the focusing structure, we can express the envelope functions $X(s)$ and $Y(s)$ as
\begin{equation}\label{eq2}
 X(s) = \bar X(s) \cdot (1+\delta_x(s)),\qquad Y(s) = \bar Y(s) \cdot (1+\delta_y(s))
\end{equation}
where
\begin{displaymath}
 |\delta_x(s)|\ll 1,\qquad |\delta_y(s)|\ll 1,
\end{displaymath}
and
\begin{displaymath}
 \delta_x(s+S)=\delta_x(s),\qquad \delta_y(s+S)=\delta_y(s).
\end{displaymath}
In the smooth approximation, the periodic lattice is replaced by equivalent continuous focusing chosen to reproduce the zero-current phase advance.
Neglecting higher-order ripple corrections gives the equations for the slowly varying envelope radii $\bar X(s)$ and $\bar Y(s)$
\begin{subequations}\label{eq3}
\begin{eqnarray}
\bar X^{\prime\prime}+k_0{}^2\cdot \bar X-2K/(\bar X+\bar Y)-\epsilon^2/\bar X^3&=&0,\label{eq3a}\\
\bar Y^{\prime\prime}+k_0{}^2\cdot \bar Y-2K/(\bar X+\bar Y)-\epsilon^2/\bar Y^3&=&0,\label{eq3b}
\end{eqnarray}
\end{subequations}
where $k_0=\sigma_0/S$.
The effective focusing strength $k_0^2$ is not the arithmetic period average of either signed focusing coefficient $\kappa_x{}^2(s)$ or $\kappa_y{}^2(s)$.
These averages can vanish in an alternating-gradient lattice despite nonzero effective focusing.
The equations remain nonlinear, but their linearization about a matched equilibrium has constant coefficients and can be solved analytically.
Angles in relations such as $k_0=\sigma_0/S$ are measured in radians, although numerical phase advances are usually quoted in degrees.
For a matched beam in this approximation, the mean transverse envelope radii are constant and equal because the effective focusing strengths and the emittances in the two directions are equal, i.e.
\begin{displaymath}
 \bar X(s) = \bar Y(s) = R = \mathrm{const.}
\end{displaymath}
Since $R^{\prime\prime}=0$ in this case, Eqs.~(\ref{eq3}) yield an algebraic equation for $R$
\begin{equation}
 R\cdot \sigma_0{}^2/S^2-K/R-\epsilon^2/R^3=0.\label{eq4}
\end{equation}
The expression $\sigma_0{}^2/S^2$, which represents the effective continuous-focusing strength, can be combined with the term $K/R^2$ representing the space-charge forces if we introduce the space-charge depressed particle tune $\sigma$ by
\begin{equation}
 \sigma^2/S^2=\sigma_0{}^2/S^2-K/R^2.\label{eq5}
\end{equation}
By substituting (\ref{eq5}) in (\ref{eq4}) one obtains for the average beam radius $R$ in the presence of space charge
\begin{equation}
R=\sqrt{\epsilon S/\sigma}. \label{eq6}
\end{equation}
For a mismatched beam, we have to solve the nonlinear system~(\ref{eq3}).
If we assume the mismatch to be small, we can linearize it.
We define the mismatch functions $x(s)$ and $y(s)$ as
\begin{equation}
\bar X(s) = R + x(s), \qquad \bar Y(s) = R + y(s),\label{eq7}
\end{equation}
where $|x|,|y|\ll R$.
By substituting (\ref{eq7}) into system (\ref{eq3}), Taylor expanding and keeping only linear terms, we get the linear differential equation
\begin{subequations}\label{eq8}
\begin{equation}
x^{\prime\prime}+x\cdot(3\sigma_0{}^2+5\sigma^2)/(2S^2) + y\cdot(\sigma_0{}^2-\sigma^2)/(2S^2) = 0,\label{eq8a}
\end{equation}
where $K/R^2=(\sigma_0{}^2-\sigma^2)/S^2$ and $\epsilon^2/R^4=\sigma^2/S^2$ have been used, as follows from Eqs.~(\ref{eq4}) and~(\ref{eq5}).
For the $y$ direction we obtain in the same way
\begin{equation}
y^{\prime\prime}+y\cdot(3\sigma_0{}^2+5\sigma^2)/(2S^2) + x\cdot(\sigma_0{}^2-\sigma^2)/(2S^2) = 0,\label{eq8b}
\end{equation}
\end{subequations}
The coupled system~(\ref{eq8}) is most easily solved by defining
\begin{equation}
z_1(s) = x(s) - y(s)\quad{\rm and}\quad z_2(s) = x(s) + y(s).
\end{equation}
Subtracting Eq.~(\ref{eq8b}) from (\ref{eq8a}), we obtain
\begin{subequations}\label{eq10}
\begin{equation}
z_1{}^{\prime\prime}+k_1{}^2\cdot z_1 = 0\quad {\rm with}\quad k_1{}^2=(\sigma_0{}^2+3\sigma^2)/S^2,\label{eq10a}
\end{equation}
whereas summing Eqs.~(\ref{eq8a}) and (\ref{eq8b}) yields
\begin{equation}
z_2{}^{\prime\prime}+k_2{}^2\cdot z_2 = 0\quad {\rm with}\quad k_2{}^2=(2\sigma_0{}^2+2\sigma^2)/S^2.\label{eq10b}
\end{equation}
\end{subequations}
The coupled system (\ref{eq8}) has been transformed into two decoupled second-order equations, which are of the same type as for the undamped harmonic oscillator.
The solutions are
\begin{subequations}\label{eq11}
\begin{eqnarray}
\left(\begin{array}{c}x(s)-y(s)\\ x^\prime(s)-y^\prime(s)\end{array}\right)
&=&\left(\begin{array}{rl}\cos k_1s & (\sin k_1s)/k_1\\ -k_1\sin k_1s & \,\,\cos k_1s\end{array}\right)
\left(\begin{array}{c}x_0-y_0\\ x_0{}^\prime-y_0{}^\prime\end{array}\right)\label{eq11a}\\
\left(\begin{array}{c}x(s)+y(s)\\ x^\prime(s)+y^\prime(s)\end{array}\right)
&=&\left(\begin{array}{rl}\cos k_2s & (\sin k_2s)/k_2\\ -k_2\sin k_2s & \,\,\cos k_2s\end{array}\right)
\left(\begin{array}{c}x_0+y_0\\ x_0{}^\prime+y_0{}^\prime\end{array}\right)\label{eq11b}
\end{eqnarray}
\end{subequations}
The oscillations in the two transverse directions described by (\ref{eq11a}) are antiparallel (``$180^\circ$ out-of-phase''), whereas those in (\ref{eq11b}) are parallel (``in-phase'').
An arbitrary linearized envelope oscillation can be expressed as a superposition of these two fundamental modes.
A pure in-phase mode requires both $x_0=y_0$ and $x_0{}^\prime=y_0{}^\prime$, whereas a pure out-of-phase mode requires both $x_0=-y_0$ and $x_0{}^\prime=-y_0{}^\prime$.
As an example, we have for the special case $x_0{}^\prime = y_0 = y_0{}^\prime=0$ the solution
\begin{subequations}\label{eq12}
\begin{eqnarray}
x(s)&=&x_0\cdot\cos\frac{1}{2}(k_1-k_2)s\cdot\cos\frac{1}{2}(k_1+k_2)s,\label{eq12a}\\
y(s)&=&x_0\cdot\sin\frac{1}{2}(k_1-k_2)s\cdot\sin\frac{1}{2}(k_1+k_2)s.\label{eq12b}
\end{eqnarray}
\end{subequations}
The envelope oscillations of the mismatched beam are thus characterized in this case by a fast spatial frequency $\frac{1}{2}(k_1+k_2)$ and a slow modulation with spatial frequency $\frac{1}{2}|k_1-k_2|$.

The spatial frequencies $k_1$ and $k_2$ can be expressed as phase shifts $\phi=kS$ of the two envelope-oscillation modes
\begin{equation}
 \phi_1 = k_1\cdot S = \sqrt{\sigma_0{}^2+3\sigma^2},\quad\phi_2 = k_2\cdot S = \sqrt{2\sigma_0{}^2+2\sigma^2}.\label{eq13}
\end{equation}
For $\sigma=\sigma_0$, i.e., zero current, the frequencies or phase shifts of the two envelope oscillations are equal, i.e.,
\begin{equation}
 \phi_1 = \phi_2 = 2\sigma_0,\label{eq14}
\end{equation}
whereas in the formal strong-space-charge limit $\sigma \to 0$ at fixed $\sigma_0$, approached by increasing the current or decreasing the emittance, the phase shifts approach the values
\begin{displaymath}
\phi_1 \to \sigma_0,\quad \phi_2 \to \sqrt{2}\sigma_0.
\end{displaymath}
\section{GENERAL LINEAR THEORY OF ENVELOPE OSCILLATIONS\\ AND INSTABILITIES\label{sec3}}
\begin{figure}[htb]
\epsfig{file=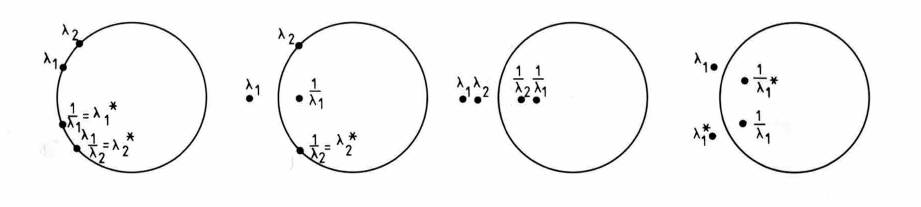,height=28mm}
\caption{Representative locations of eigenvalues $\lambda_i$ for envelope perturbations. The real pairs shown are negative, as in the half-integer resonances discussed in the text.}
\label{fig2}
\end{figure}

Since the smooth approximation cannot describe the instability bands that occur for $\sigma_0$ greater than about $90^\circ$, it is necessary to retain the periodic focusing coefficients in the linear analysis of envelope perturbations.
Abandoning the use of ``smoothed'', i.e., averaged quantities such as the mean beam radius $R$, we arrive at a system of coupled linear differential equations with periodic rather than constant coefficients, which we solve numerically.
The starting point is again the nonlinear system (\ref{eq1}), in which we substitute the perturbed envelope functions directly
\begin{equation}
 X(s)=X_0(s)+x(s), \quad Y(s)=Y_0(s)+y(s).\label{eq15}
\end{equation}
Here $X_0$ and $Y_0$ denote the matched envelope functions, i.e., periodic solution functions of (\ref{eq1}) and $x$, $y$ denote small perturbations
\begin{displaymath}
 |x(s)| \ll X_0(s), \quad |y(s)| \ll Y_0(s).
\end{displaymath}
Due to these conditions, we can linearize the differential equations for the perturbation functions $x(s)$ and $y(s)$ and obtain
\begin{subequations}\label{eq16}
\begin{eqnarray}
x^{\prime\prime}(s)+a_1(s) \cdot x(s)+a_0(s) \cdot y(s)&=&0,\label{eq16a}\\
y^{\prime\prime}(s)+a_2(s) \cdot y(s)+a_0(s) \cdot x(s)&=&0,\label{eq16b}
\end{eqnarray}
\end{subequations}
with three $S$-periodic coefficients
\begin{eqnarray*}
 a_0(s)&=&2K/{\big(X_0(s)+Y_0(s)\big)}^2,\\
 a_1(s)&=&\kappa_x{}^2(s)+3\epsilon^2/X_0{}^4(s)+a_0(s),\\
 a_2(s)&=&\kappa_y{}^2(s)+3\epsilon^2/Y_0{}^4(s)\,+a_0(s).
\end{eqnarray*}
To solve this system, we need the matched envelope functions $X_0(s)$ and $Y_0(s)$.
The two second-order equations (\ref{eq16}) are equivalent to a system of four first-order differential equations.
With $z=(x,x^\prime,y,y^\prime)^T$, we may write in matrix notation
\begin{equation}
 z^\prime(s)=A(s)\cdot z(s),\label{eq17}
\end{equation}
with the $S$-periodic matrix
\begin{displaymath}
 A(s)=\left(
\begin{array}{cccc}0 & 1 & 0 & 0\\
 -a_1(s) & 0 & -a_0(s) & 0\\
 0 & 0 & 0 & 1\\
 -a_0(s) & 0 & -a_2(s) & 0
\end{array}
 \right).
\end{displaymath}
If $Z(s)$ denotes the $4 \times 4$ solution matrix of (\ref{eq17}) with $Z(0)=E~(E=$ unit matrix), we may write Floquet's theorem as
\begin{equation}
Z(s+nS) = Z(s) \cdot Z(S)^n,\label{eq18}
\end{equation}
where $n$ is an arbitrary integer number.
Equation~(\ref{eq18}) expresses propagation over successive periods in terms of the one-period matrix $Z(S)$ and propagation within a reference period.
If $Z(S)$ has a complete set of eigenvectors, we obtain a nonsingular $4 \times 4$ matrix of eigenvectors $C$ and a diagonal matrix of eigenvalues (denoted by $\Lambda$)
\begin{displaymath}
 Z(S)\cdot C=C \cdot \Lambda.
\end{displaymath}
With the matrix $Y(s)$ defined by
\begin{displaymath}
 Y(s)= Z(s) \cdot C,
\end{displaymath}
it follows from (\ref{eq18}) that
\begin{equation}
Y(s+nS)=Y(s)\cdot\Lambda^n.\label{eq19}
\end{equation}
Under this assumption, every solution $z(s)$ of (\ref{eq17}) can be expressed as a linear combination of the columns of $Y(s)$.
All perturbations remain bounded only if the powers of the one-period matrix remain bounded.
To establish its symplecticity, introduce
\begin{displaymath}
 J=\left(\begin{array}{rrrr}
 0&1&0&0\\ -1&0&0&0\\ 0&0&0&1\\ 0&0&-1&0
 \end{array}\right).
\end{displaymath}
The matrix $A(s)$ satisfies $A^T J+JA=0$, and hence
\begin{displaymath}
 \frac{d}{ds}\bigl(Z^T JZ\bigr)=Z^T(A^T J+JA)Z=0.
\end{displaymath}
Since $Z(0)=E$, it follows that $Z(s)^T JZ(s)=J$.
Thus $Z(s)$ is real and symplectic, so its eigenvalues occur in reciprocal and complex-conjugate pairs.
Boundedness for all initial perturbations requires that all eigenvalues lie on the unit circle and that $Z(S)$ be diagonalizable.
For distinct eigenvalues, the latter condition is automatic.
At repeated eigenvalues, nontrivial Jordan blocks can produce polynomial growth even if all eigenvalues have unit modulus.
If we express the eigenvalues in polar coordinates,
\begin{equation}
\lambda=|\lambda|\cdot e^{i\phi},\label{eq20}
\end{equation}
we obtain the eigenvalue configurations illustrated in Fig.~\ref{fig2} for the cases discussed below.\cite{courant}

By means of (\ref{eq20}), we distinguish the amplitude multiplier $|\lambda|$ per focusing period from the phase shift $\phi$ of the corresponding envelope oscillation.
The exponential growth rate per unit length of a growing mode is
\begin{displaymath}
 \Gamma=\frac{\ln|\lambda|}{S}.
\end{displaymath}
An eigenvalue with $|\lambda|>1$ indicates exponential instability.
A reciprocal multiplier smaller than unity describes a contracting solution direction, not dissipative damping.
In the instability bands studied below, the following resonance configurations occur.
\begin{enumerate}
\item[(a)]~~~One or both reciprocal eigenvalue pairs lie on the negative real axis, so the corresponding phase shift is $180^\circ$.
\item[(b)]~~~The eigenvalues form a nonreal reciprocal-conjugate quartet off the unit circle, with correlated phase branches satisfying $\frac{1}{2}(\phi_1+\phi_2)=180^\circ$ in the convention used in the plots.
\end{enumerate}
Case (a) is associated with a half-integer resonance between the focusing structure and an envelope-oscillation mode (``parametric'' resonance).
Case (b) is associated with the coupling of the two envelope-oscillation modes (``confluent'' resonance).
These phase relations characterize the resonance configurations but are not, by themselves, sufficient conditions for instability.
A general real symplectic matrix can also have positive real reciprocal eigenvalues, whose phase is $0^\circ$ modulo $360^\circ$.

\begin{figure}[H]
\hspace*{-2mm}\centering\epsfig{file=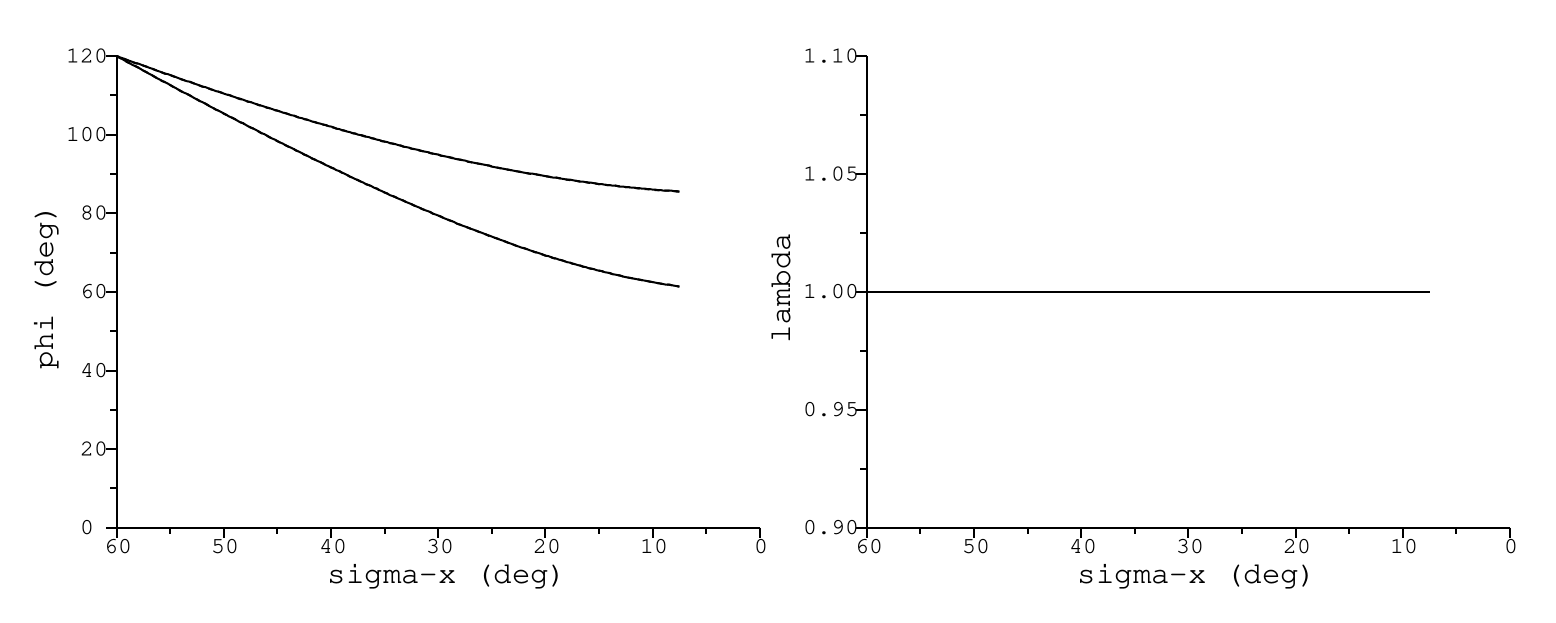,width=0.3\linewidth,angle=-90}
\hspace*{-2mm}\centering\epsfig{file=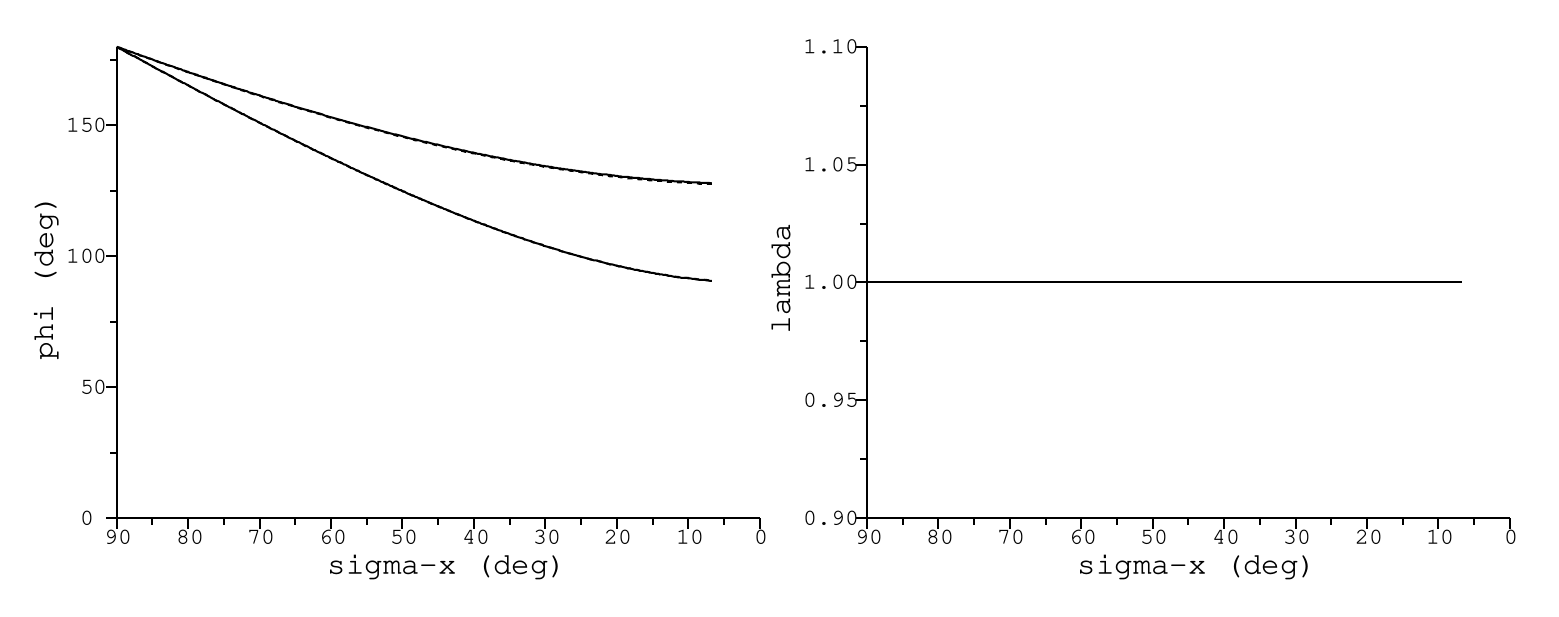,width=0.3\linewidth,angle=-90}
\hspace*{-3mm}\centering\epsfig{file=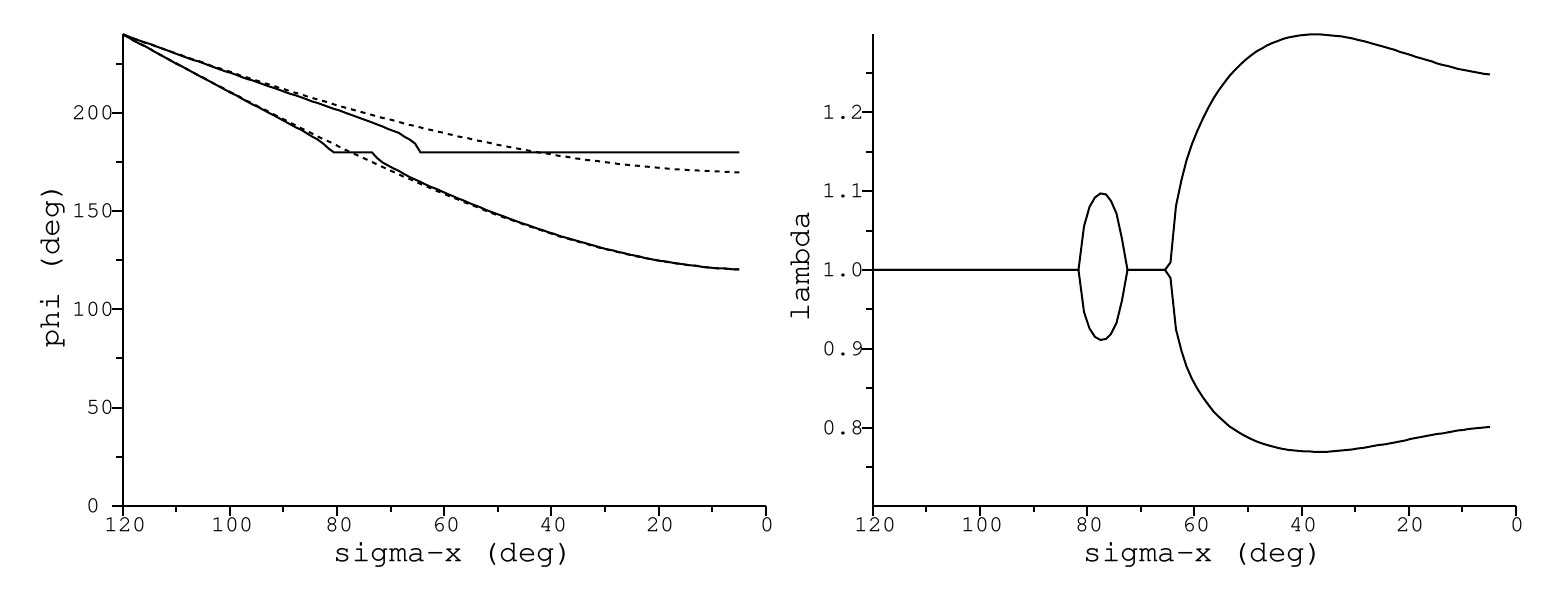,width=0.3\linewidth,angle=-90}
\hspace*{-6mm}\centering\epsfig{file=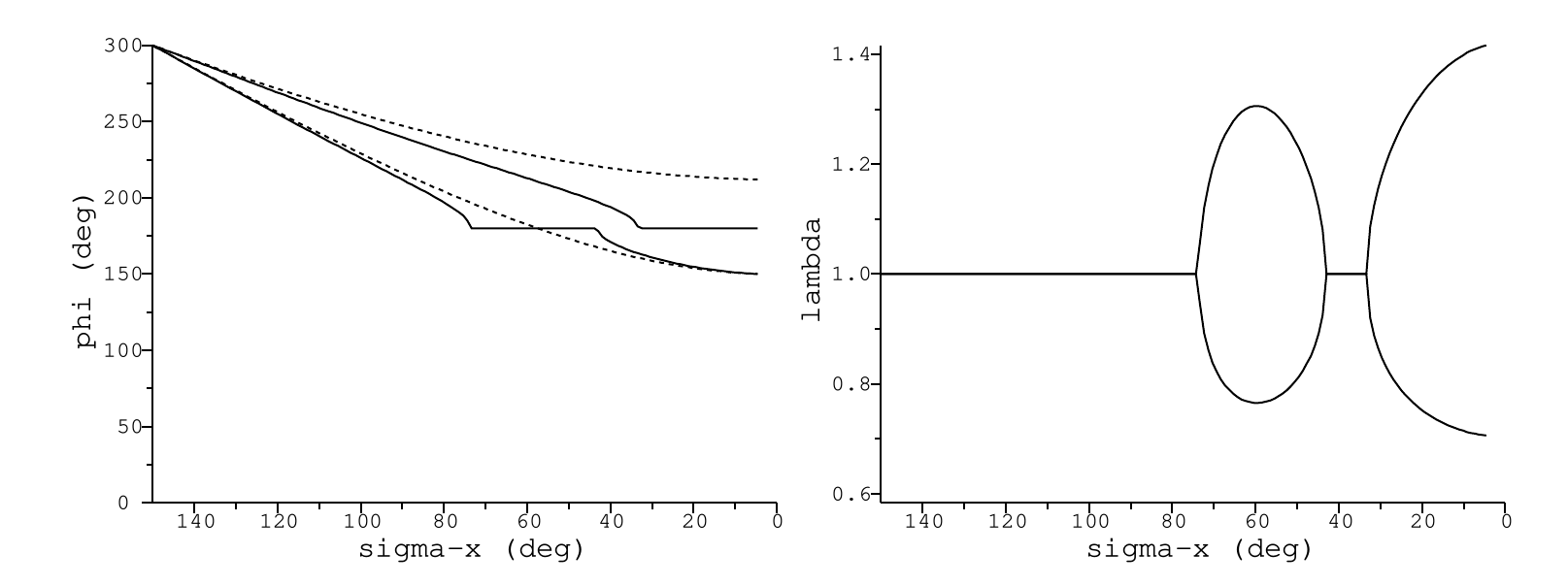,width=0.3\linewidth,angle=-90}
\caption{Phase shifts and growth factors per period of envelope perturbations versus $\sigma$ for $\sigma_0=60^\circ, 90^\circ, 120^\circ, 150^\circ$ for the Maryland solenoid channel. (Dotted curves represent the smooth approximation results.)}
\label{fig3}
\end{figure}

\begin{figure}[H]
\hspace*{1mm}\centering\epsfig{file=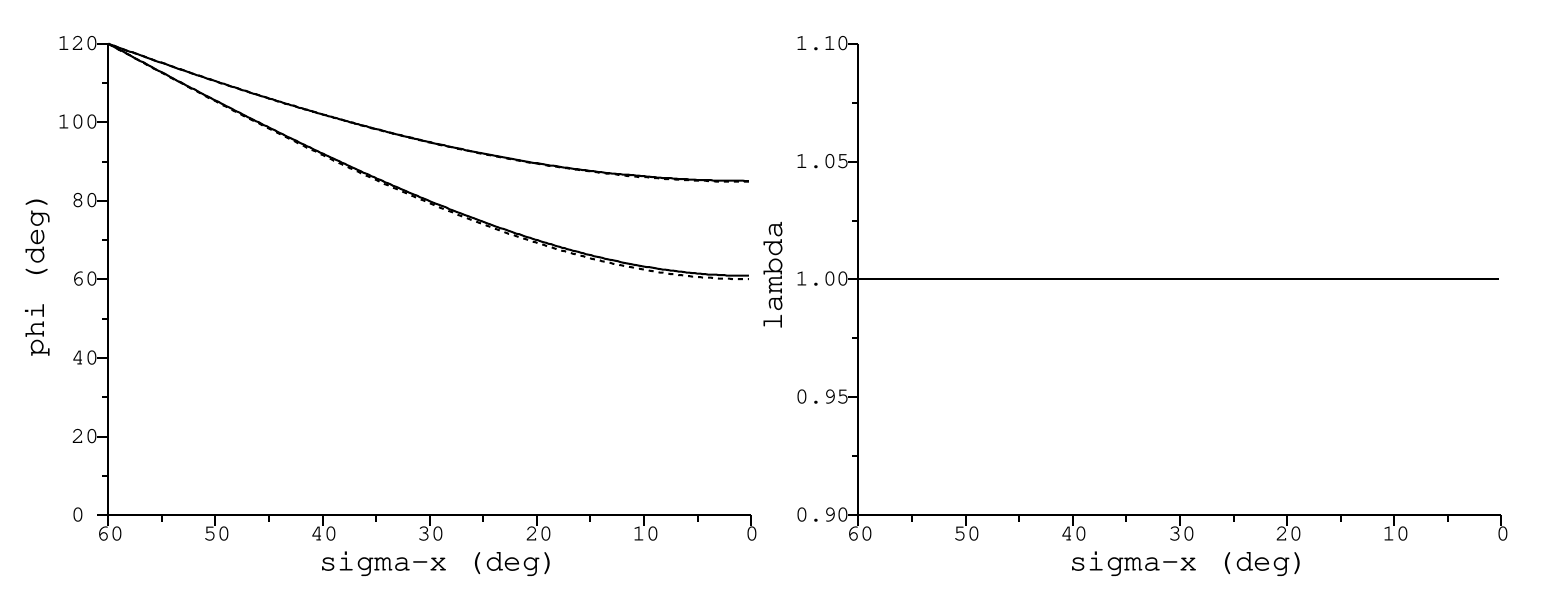,width=0.3\linewidth,angle=-90}
\centering\epsfig{file=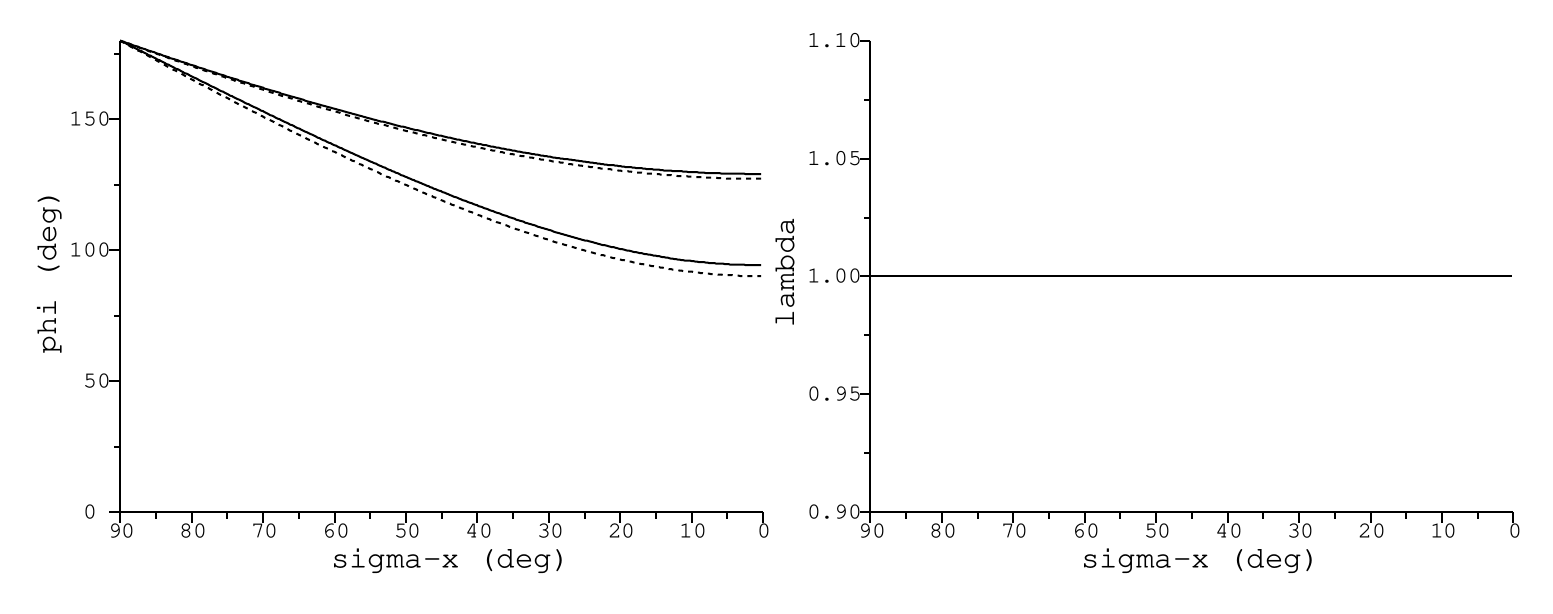,width=0.3\linewidth,angle=-90}
\centering\epsfig{file=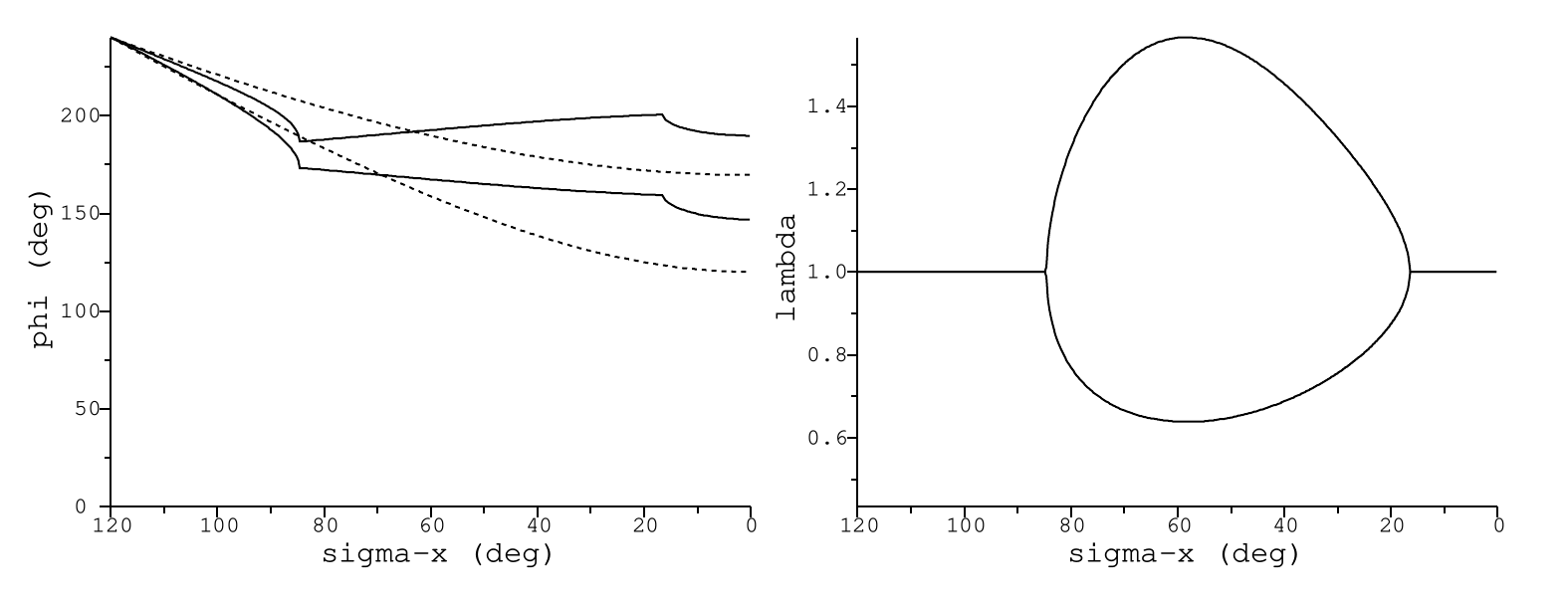,width=0.3\linewidth,angle=-90}
\centering\epsfig{file=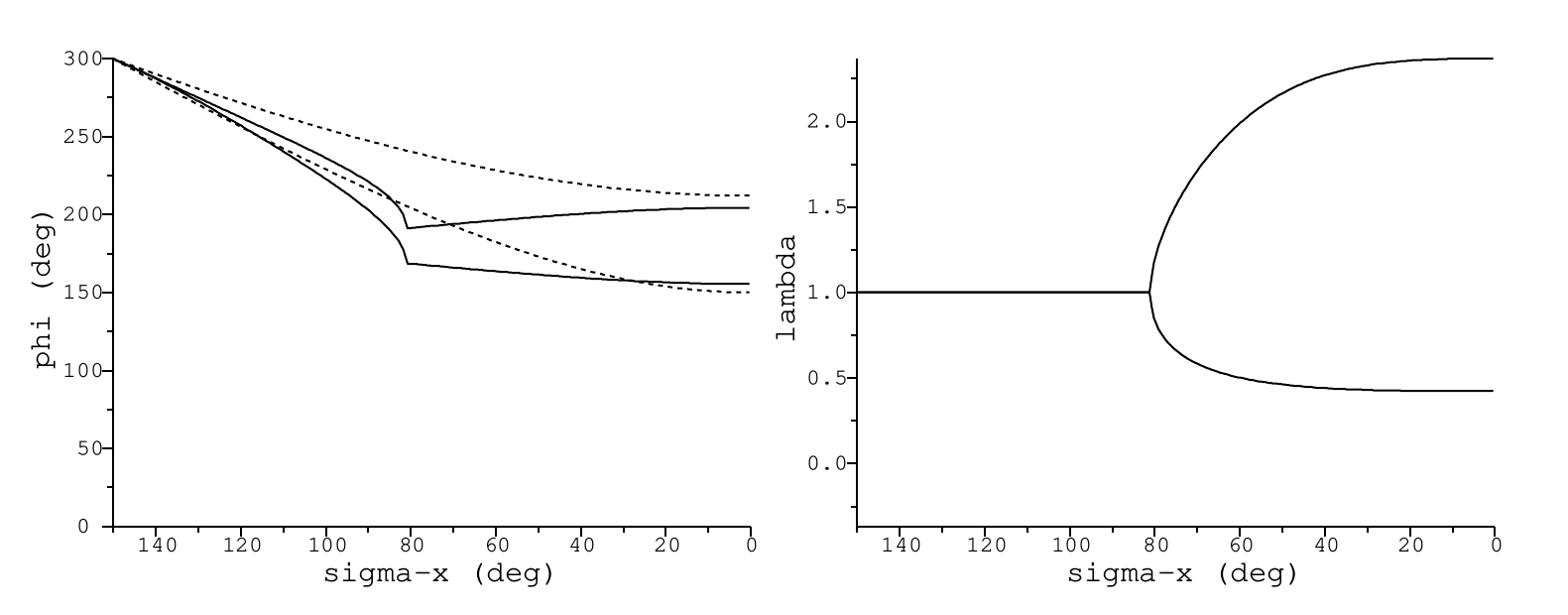,width=0.3\linewidth,angle=-90}
\caption{Phase shifts and growth factors per period of envelope perturbations versus $\sigma$ for $\sigma_0=60^\circ, 90^\circ, 120^\circ, 150^\circ$ for the GSI quadrupole channel. (Dotted curves represent the smooth approximation results.)}
\label{fig4}
\end{figure}

The results of numerical integrations of Eq.~(\ref{eq17}) are plotted in Fig.~\ref{fig3} using the geometry and beam parameters of the Maryland solenoid channel and in Fig.~\ref{fig4} for the GSI quadrupole channel.
The left part of the appropriate figure shows the $\phi_1$, $\phi_2$-values versus $\sigma$ for a fixed $\sigma_0$, the right part the amplitude multipliers $|\lambda|$ per focusing period versus $\sigma$.
Instability is indicated by $|\lambda|$-values differing from unity.
The solid $\phi$-lines are the perturbation phase shifts obtained by numerical integration and eigenvalue analysis.
The dotted lines show the results obtained from the smooth-approximation theory~(\ref{eq13}).
For $\sigma_0=60^\circ$ and $\sigma_0=90^\circ$, these results are nearly identical, although for the quadrupole channel a slight difference can be seen in the case $\sigma_0=90^\circ$.
Above $\sigma_0=90^\circ$, instability occurs in some specific regions.

For the solenoid channel, the unstable bands shown are ``parametric'' resonances with a phase shift of $180^\circ$.
For the round matched solution $X_0=Y_0=R_0(s)$, Eqs.~(\ref{eq16}) decouple exactly in the linearized theory into
\begin{eqnarray*}
 z_1^{\prime\prime}+\left(\kappa^2+\frac{3\epsilon^2}{R_0^4}\right)z_1&=&0,\\
 z_2^{\prime\prime}+\left(\kappa^2+\frac{3\epsilon^2}{R_0^4}+\frac{K}{R_0^2}\right)z_2&=&0,
\end{eqnarray*}
where $z_1=x-y$ and $z_2=x+y$.
Thus these two linear modes cannot undergo a coupled confluent instability with each other.
Near the parametric instability bands, the smooth-approximation results differ from the periodic-coefficient results, but approach them again outside these bands.

In the case of the quadrupole channel, we encounter ``confluent'' resonances in the illustrated cases with $\sigma_0=120^\circ$ and $150^\circ$, over specific intervals of the depressed phase advance $\sigma$.
For both types of beam transport channels, the maximum instability growth factors increase significantly with $\sigma_0$ in the examples shown.

The results obtained here from the perturbation theory of the K-V envelope equations are equivalent to those obtained from the Vlasov-equation perturbation analysis\cite{hofmann} for the special case of the ``second-order even'' mode.
It should be noted here that the Vlasov approach leads to an additional $xy$ (``second-order odd'') perturbation for a FODO channel with an asymmetry.\cite{hofmann}
This $xy$ mode, which is not contained in our analysis, can result in an instability for $\sigma_0>90^\circ$ of relatively small growth rate, as was pointed out to us by L.~J.~Laslett, who graciously performed some calculations of this mode pertaining to the GSI channel.\cite{laslett}

\section{NUMERICAL SOLUTIONS OF THE FULL K-V EQUATIONS\\ FOR MISMATCHED BEAMS\label{sec4}}
\subsection{Solenoid Channel}
\begin{figure}[H]
\centering\epsfig{file=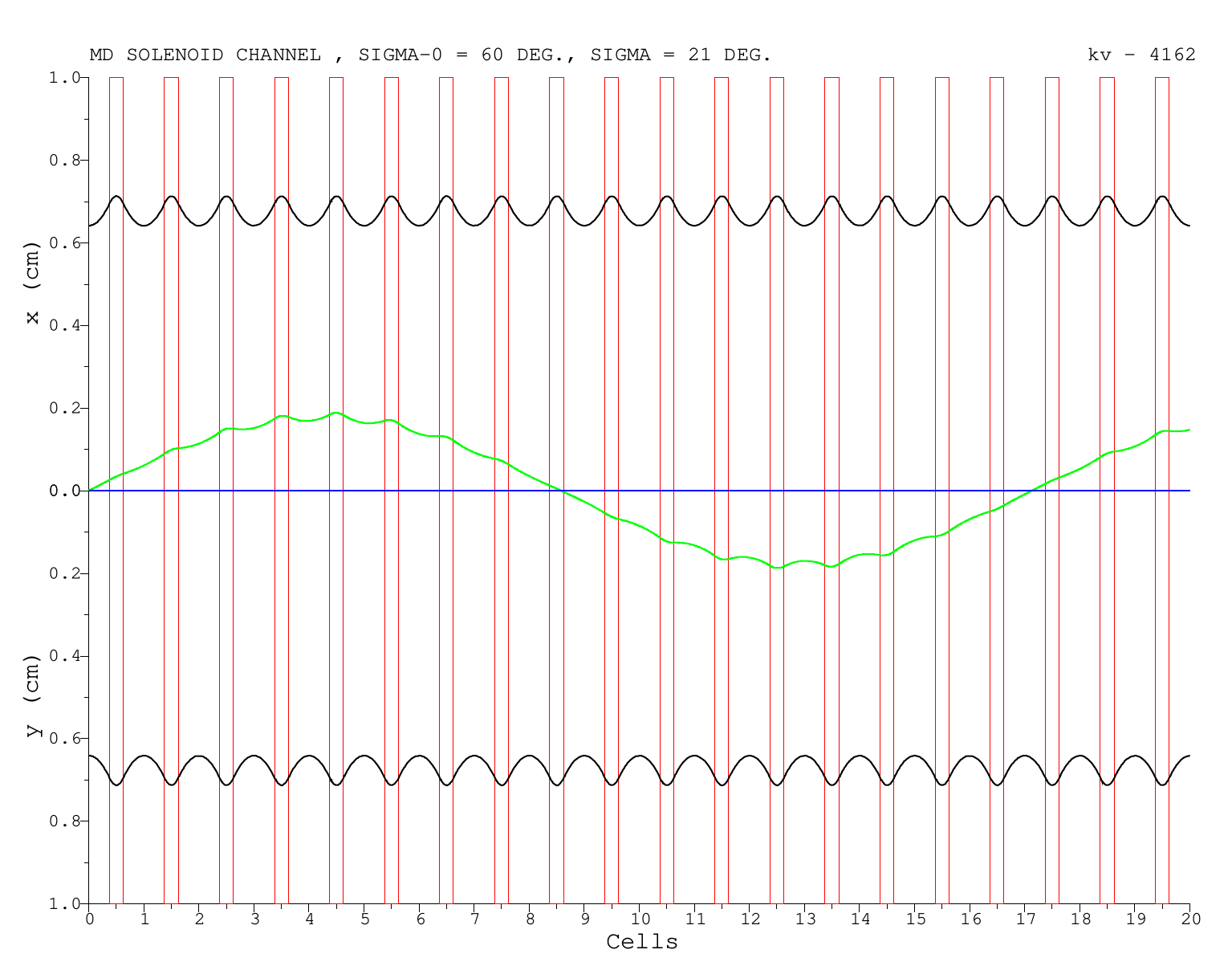,width=0.66\linewidth,angle=-90}
\caption{Solenoid channel --- matched beam ($\sigma_0=60^\circ, \sigma=21^\circ$).}
\label{fig5}
\end{figure}

\begin{figure}[H]
\centering\epsfig{file=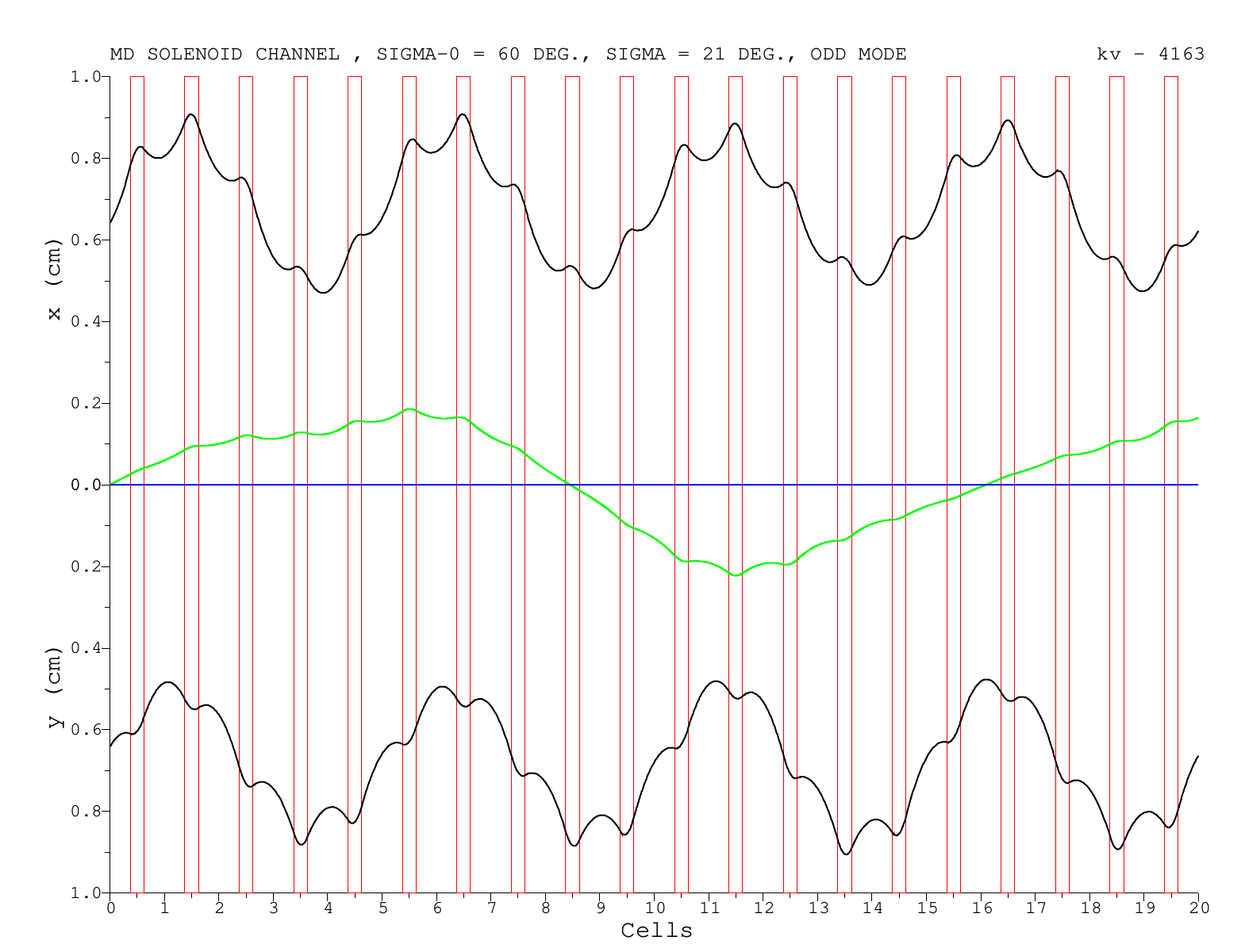,width=0.66\linewidth,angle=-90}
\caption{Solenoid channel --- ``$180^\circ$ out-of-phase'' mode ($\sigma_0=60^\circ, \sigma=21^\circ$).}
\label{fig6}
\end{figure}

\begin{figure}[H]
\centering\epsfig{file=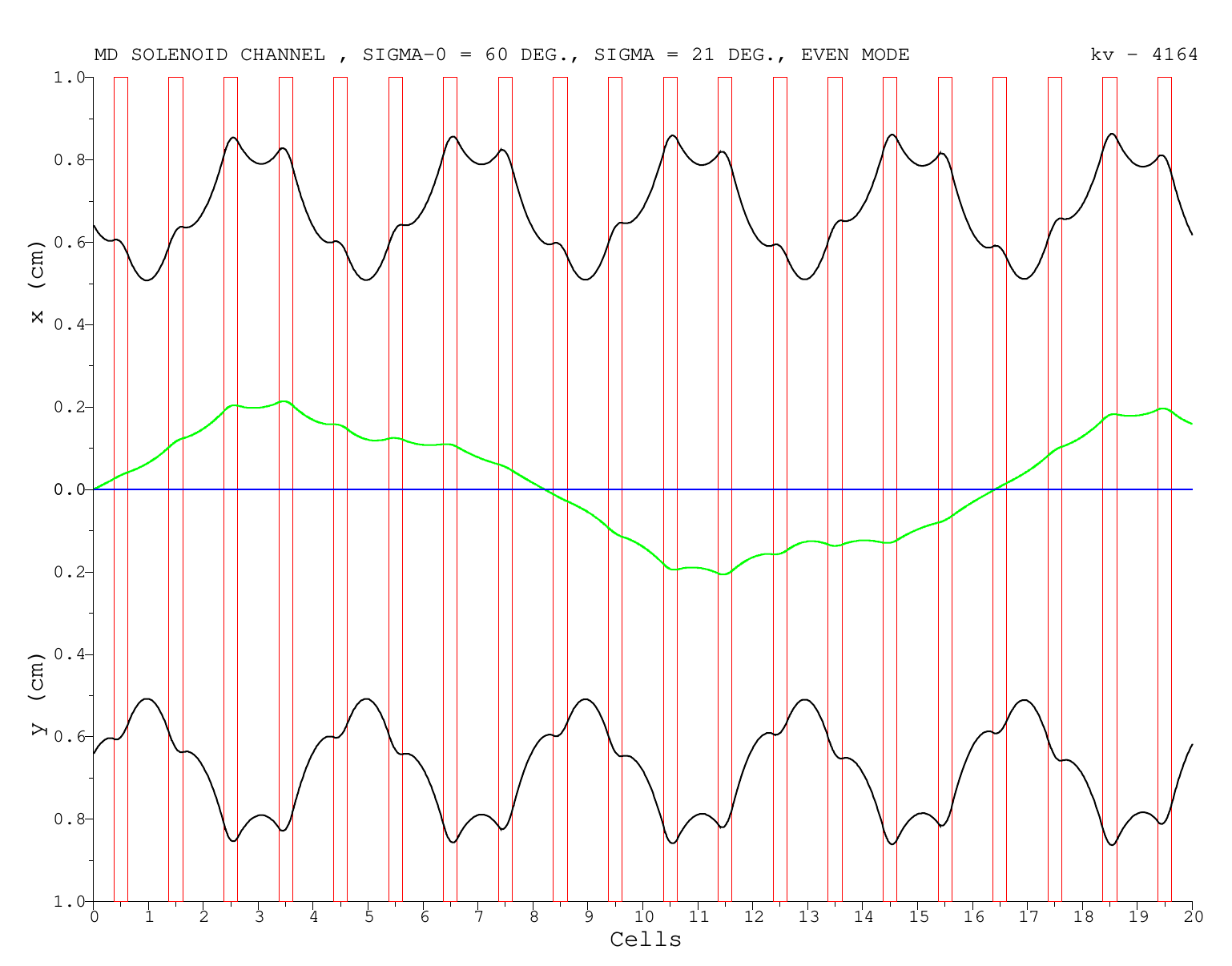,width=0.66\linewidth,angle=-90}
\caption{Solenoid channel --- ``in-phase'' mode ($\sigma_0=60^\circ, \sigma=21^\circ$).}
\label{fig7}
\end{figure}

\begin{figure}[H]
\centering\epsfig{file=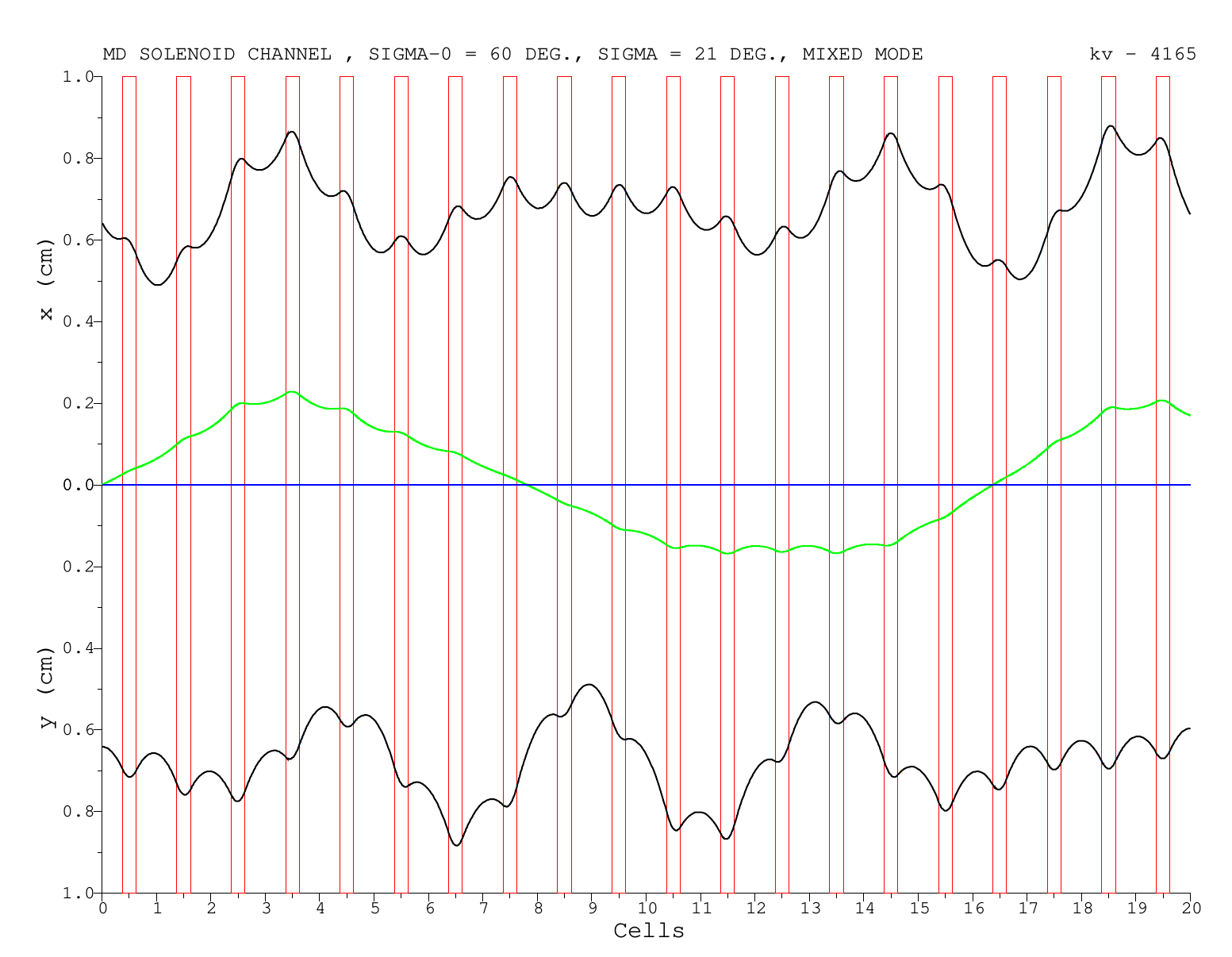,width=0.6\linewidth,angle=-90}
\caption{Solenoid channel --- ``mixed'' mode ($\sigma_0=60^\circ, \sigma=21^\circ$).}
\label{fig8}
\end{figure}

\begin{figure}[H]
\centering\epsfig{file=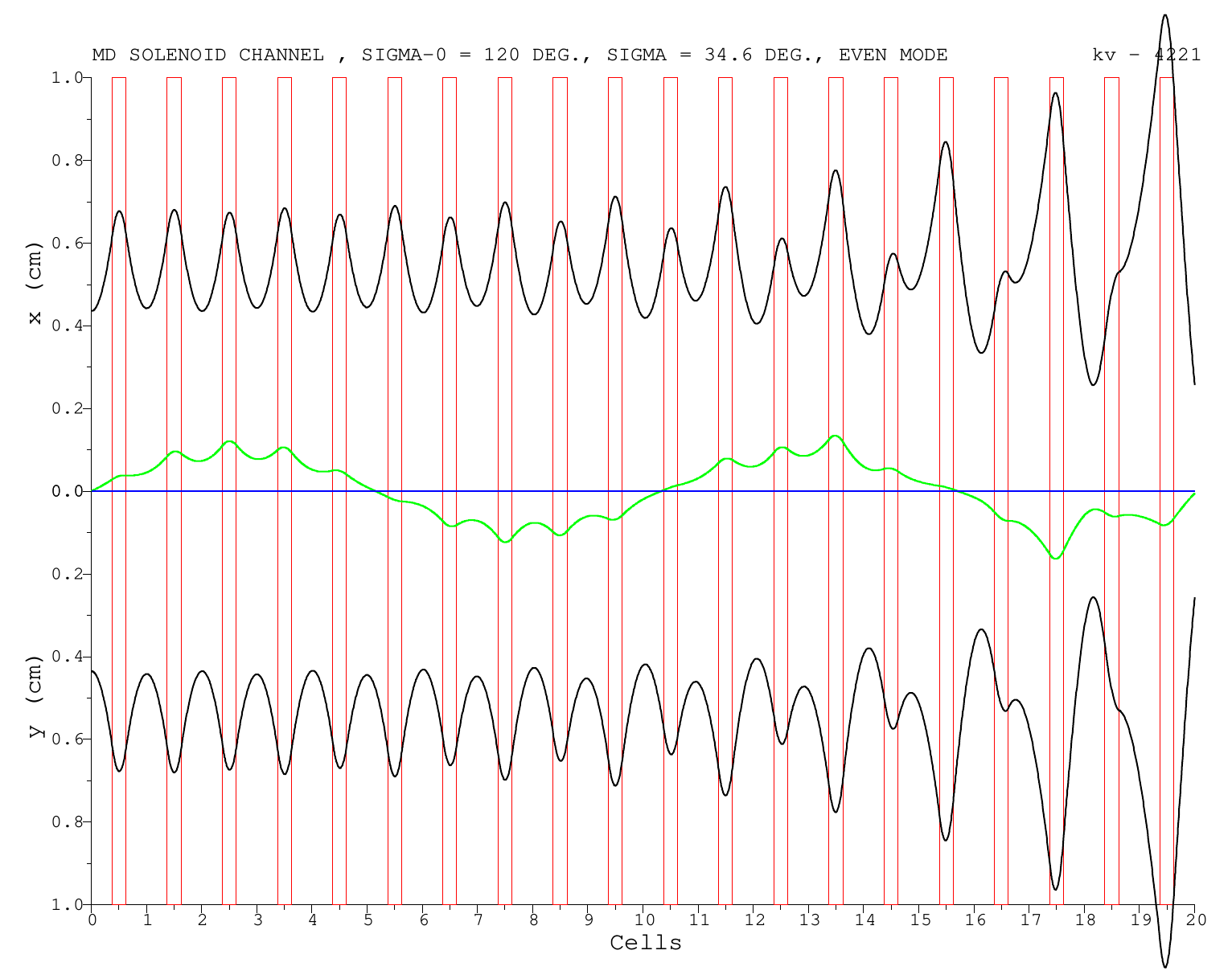,width=0.6\linewidth,angle=-90}
\caption{Solenoid channel --- unstable ``in-phase'' mode ($\sigma_0=120^\circ, \sigma=34.6^\circ$).}
\label{fig9}
\end{figure}

\begin{figure}[H]
\centering\epsfig{file=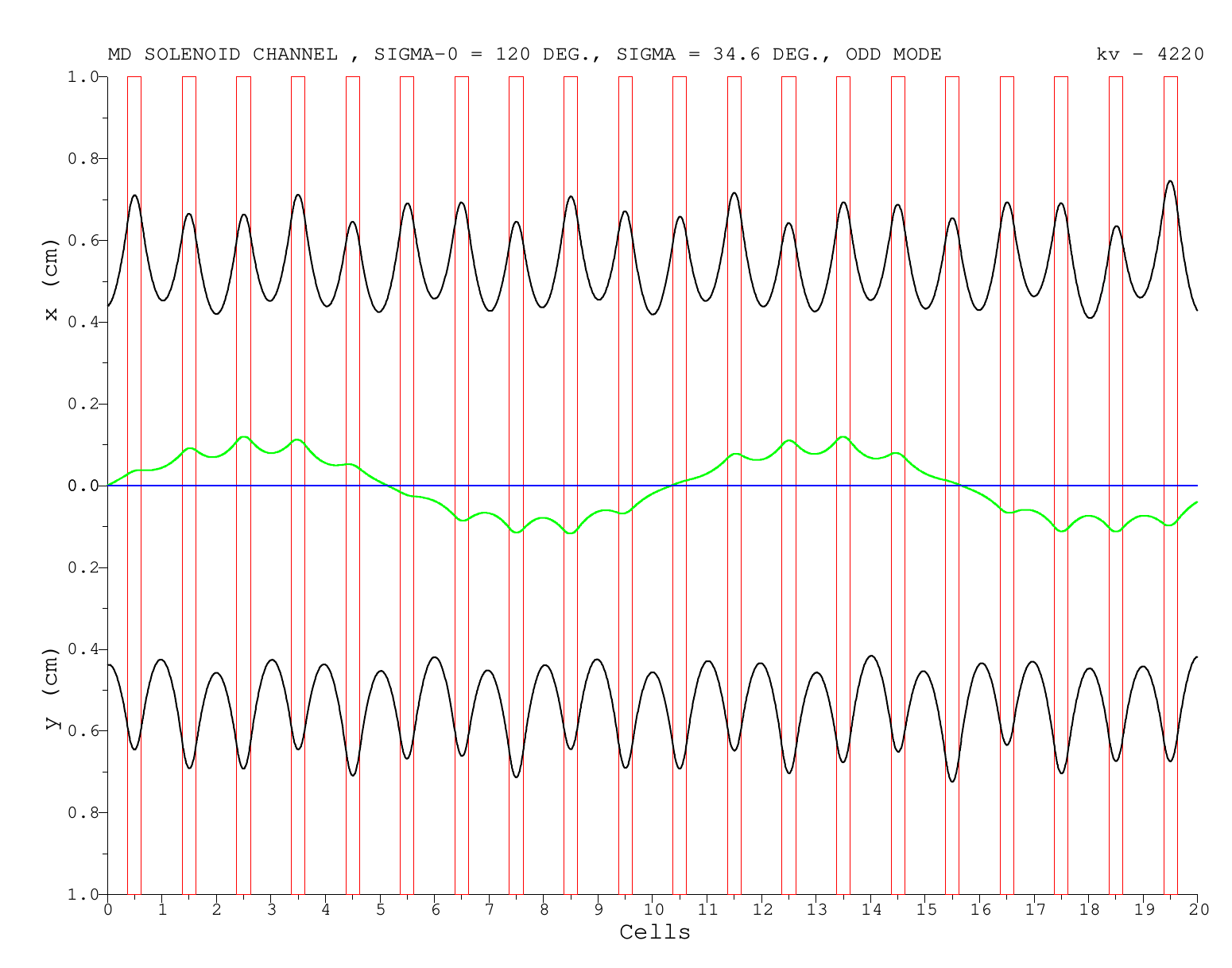,width=0.66\linewidth,angle=-90}
\caption{Solenoid channel --- ``$180^\circ$ out-of-phase'' mode, stable in the linearized theory ($\sigma_0=120^\circ, \sigma=34.6^\circ$).}
\label{fig10}
\end{figure}

As a check of our linearized envelope perturbation theory and to illustrate the beam behavior in the case of mismatch conditions, we performed numerical integration studies of the full nonlinear K-V equations for both solenoid and quadrupole channels.
The parameters of the Maryland and GSI transport channels were chosen for these studies.
However, the results are typical for the general behavior of mismatched beams in solenoid and quadrupole channels.
Figure~\ref{fig5} shows a beam which is matched to the Maryland solenoid channel for $\sigma_0=60^\circ$.
Due to the assumed beam current, the tune is depressed to $\sigma=21.2^\circ$, as can be seen from the single-particle trajectory which has a betatron oscillation wavelength of about $17$ periods.
The envelope is an exact $S$-periodic function.
It should be noted here that in Figs.~\ref{fig5} to~\ref{fig16} the particle trajectories are shown in the $y=0$ plane of the model coordinates, which are Larmor-frame coordinates for the solenoid channel.
In the linearized theory, the two pure eigenmodes are selected by the initial envelope deviations and slopes.
The finite-amplitude integrations below illustrate these mode patterns, but nonlinear corrections need not preserve exact mode purity.
Figure~\ref{fig6} shows an ``$180^\circ$ out-of-phase'' mismatch with the smooth-approximation phase shift
\begin{displaymath}
\phi_1 = \sqrt{\sigma_0{}^2+3\sigma^2}=70.3^\circ.
\end{displaymath}
Thus, the perturbation has an approximate wavelength of $360/70.3\simeq5$ structure periods, as can be seen from the envelope plot.
For an ``in-phase'' mismatch mode, the smooth approximation gives a perturbation phase shift of
\begin{displaymath}
\phi_2= \sqrt{2\sigma_0{}^2+2\sigma^2}=90^\circ,
\end{displaymath}
and therefore an approximate envelope-oscillation wavelength of $4$ periods.
This case is shown in Fig.~\ref{fig7}.

An example of a ``mixed-mode'' mismatch, which is the usual type in practice, is shown in Fig.~\ref{fig8}, where a fast envelope modulation is evident, but no periodicity, i.e., no repetition of the oscillation pattern, can be seen within the first $20$ cells.

To illustrate the two modes in an unstable parameter region with $\sigma_0>90^\circ$, we choose $\sigma_0=120^\circ$ and the space charge depressed tune of $\sigma=34.6^\circ$.
As can be seen from Fig.~\ref{fig3}, the ``in-phase'' oscillation mode is unstable due to a parametric resonance ($\phi_2=180^\circ$) with growth factor per period of $|\lambda|=1.283$, whereas the ``$180^\circ$ out-of-phase'' mode is stable ($\phi_1=134^\circ, |\lambda|=1$).
These facts are illustrated by the computer results plotted in Fig.~\ref{fig9} showing growth of the ``in-phase'' perturbation and Fig.~\ref{fig10} showing bounded, predominantly out-of-phase oscillations over the displayed interval.
The latter example does not establish stability of the matched beam against arbitrary perturbations, since an in-phase component can still grow.


\subsection{Quadrupole Channel}
\begin{figure}[H]
\centering\epsfig{file=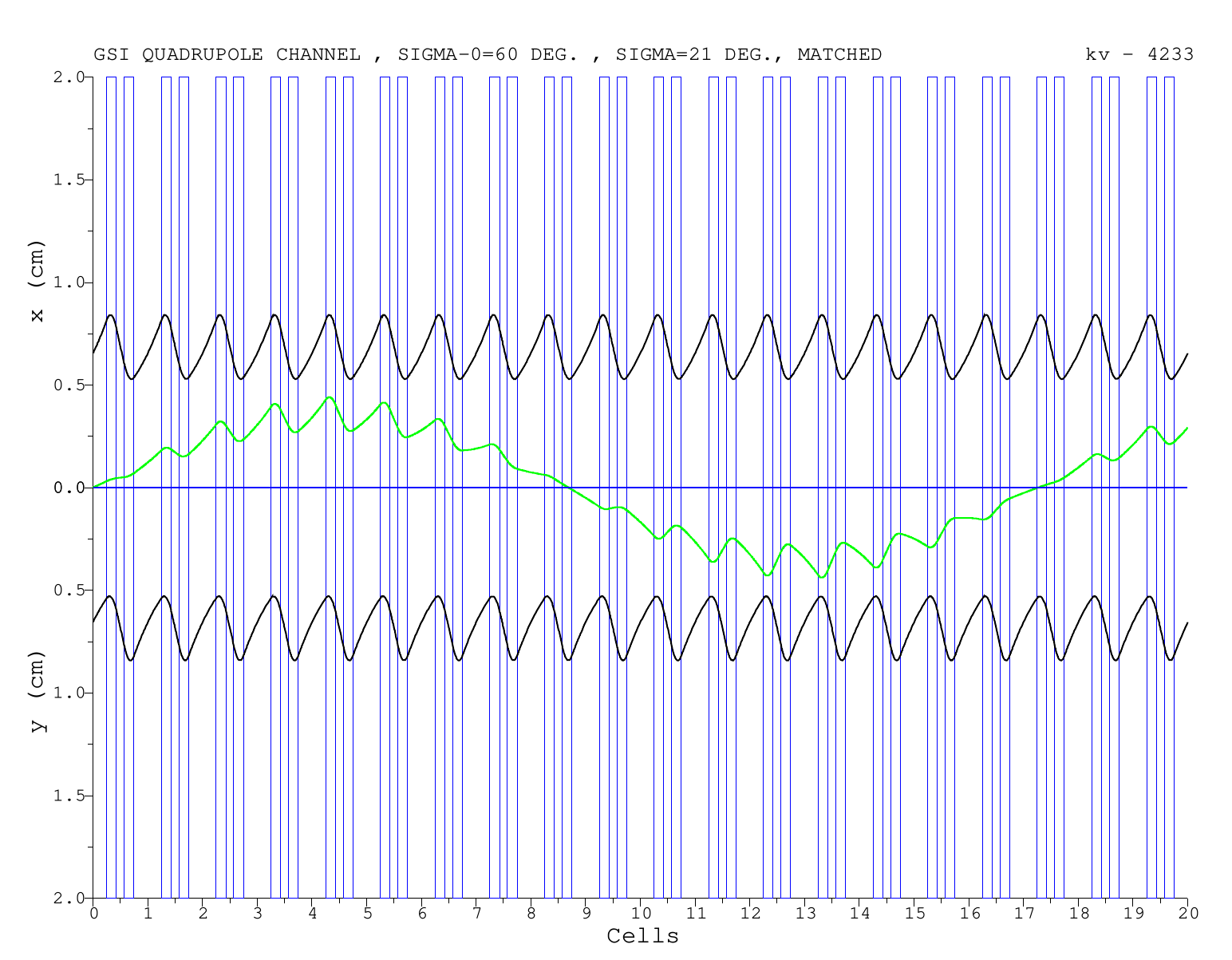,width=0.66\linewidth,angle=-90}
\caption{Quadrupole channel --- matched beam ($\sigma_0=60^\circ, \sigma=21^\circ$).}
\label{fig11}
\end{figure}

\begin{figure}[H]
\centering\epsfig{file=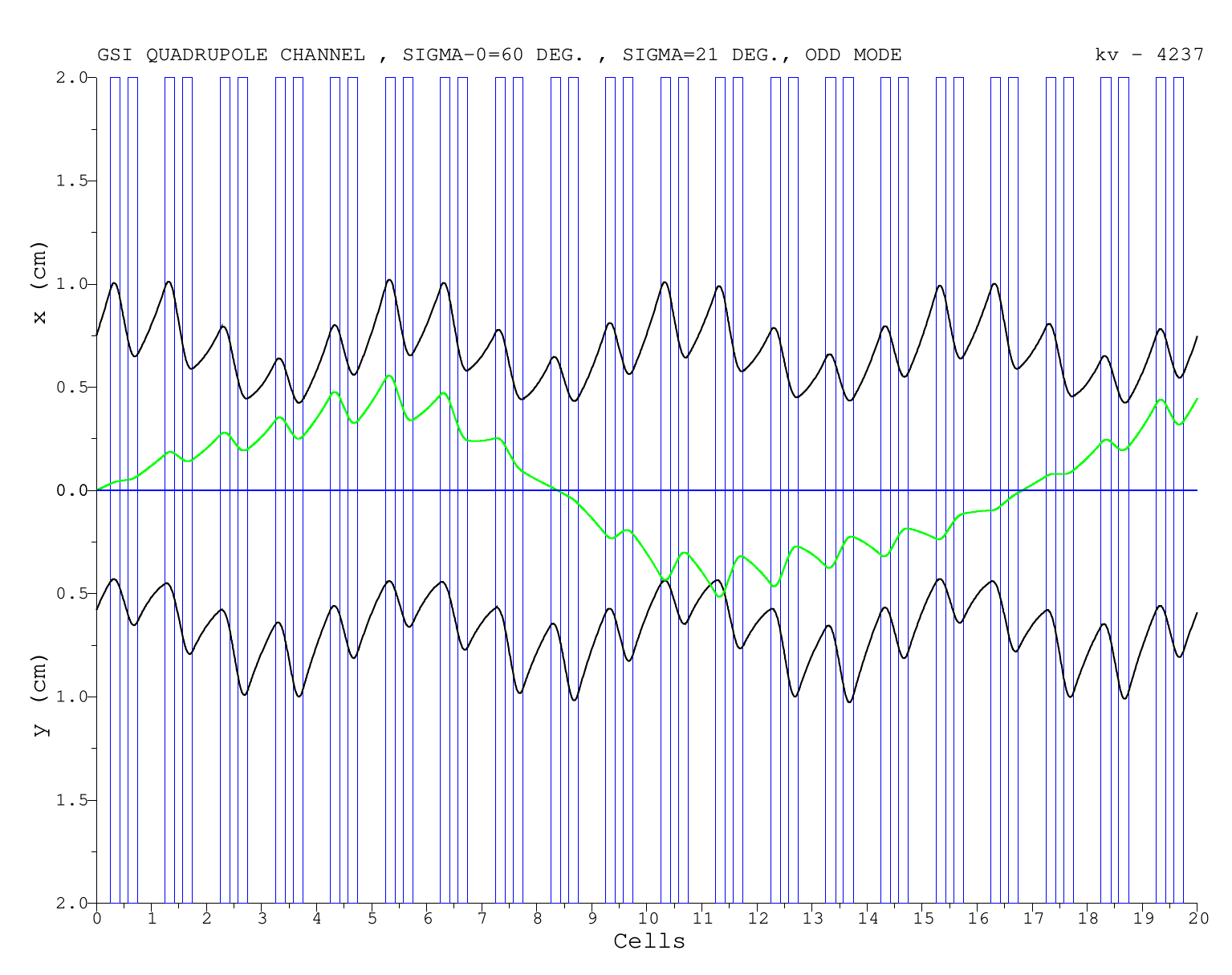,width=0.66\linewidth,angle=-90}
\caption{Quadrupole channel --- ``$180^\circ$ out-of-phase'' mode ($\sigma_0=60^\circ, \sigma=21^\circ$).}
\label{fig12}
\end{figure}

\begin{figure}[H]
\centering\epsfig{file=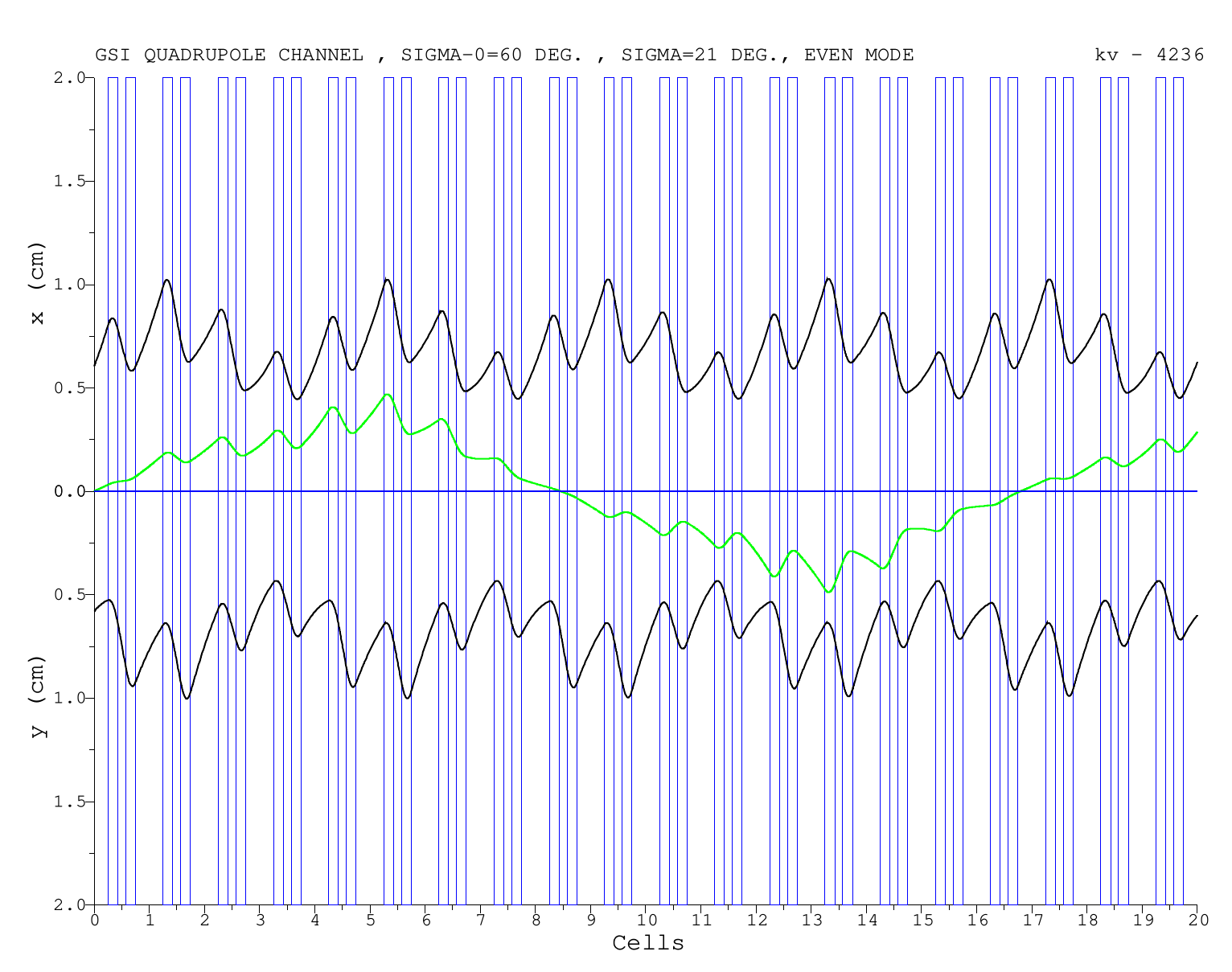,width=0.66\linewidth,angle=-90}
\caption{Quadrupole channel --- ``in-phase'' mode ($\sigma_0=60^\circ, \sigma=21^\circ$).}
\label{fig13}
\end{figure}

\begin{figure}[H]
\centering\epsfig{file=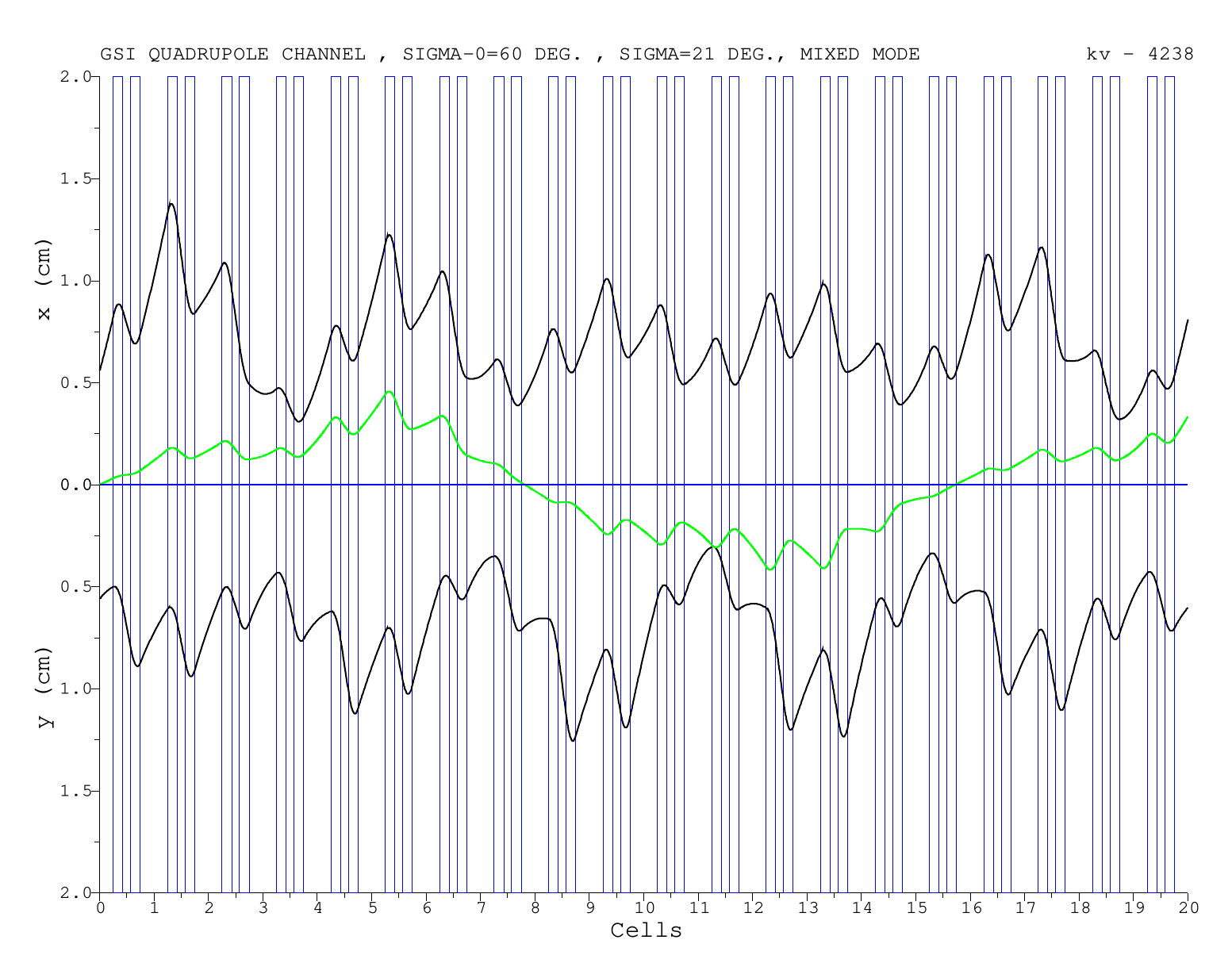,width=0.66\linewidth,angle=-90}
\caption{Quadrupole channel --- ``mixed'' beam ($\sigma_0=60^\circ, \sigma=21^\circ$).}
\label{fig14}
\end{figure}

Similar results are obtained for mismatched beams in a quadrupole channel.
As an example, we take here the data of the GSI quadrupole channel.
For $\sigma_0= 60^\circ$ and a space-charge depressed tune of $\sigma=21^\circ$, Fig.~\ref{fig11} shows the matched envelope function and, in addition, a single-particle trajectory with the betatron-oscillation wavelength of $360/21\simeq 17$ cells.
The envelope oscillations due to mismatch are shown in Fig.~\ref{fig12} for the ``$180^\circ$ out-of-phase'' mode with an approximate wavelength of $5$ cells, and in Fig.~\ref{fig13} for the ``in-phase'' mode with an approximate wavelength of $4$ cells.
For the periodic quadrupole lattice, these names denote continuations of the smooth-approximation modes, not exact pointwise relations $x(s)=\pm y(s)$ throughout a cell.
An exact pure mode of the linearized periodic system must be launched in the real invariant subspace associated with the appropriate pair of Floquet eigenvectors, using both envelope deviations and slopes.
A ``mixed'' mode is shown in Fig.~\ref{fig14}.

\begin{figure}[H]
\centering\epsfig{file=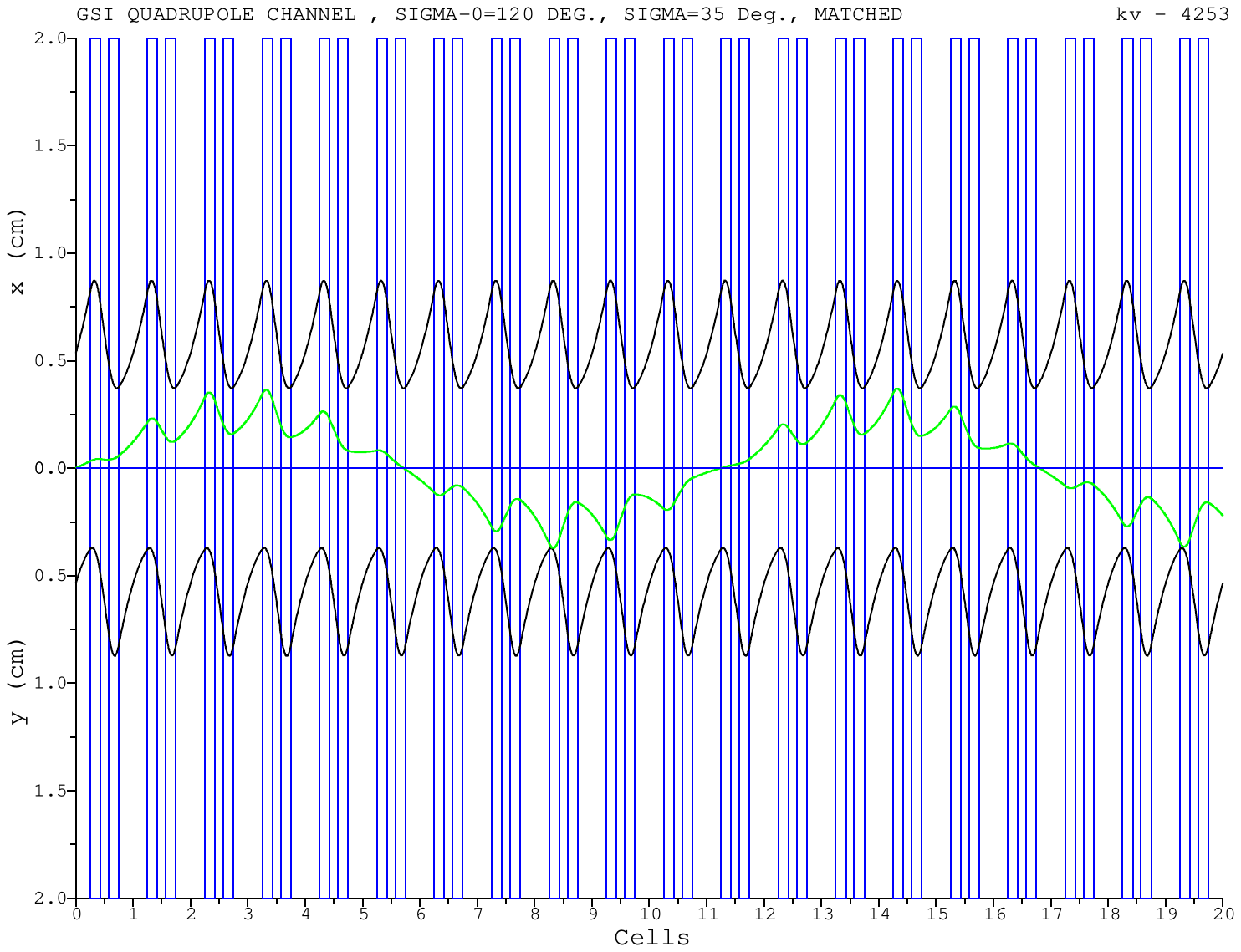,width=0.66\linewidth,angle=-90}
\caption{Quadrupole channel --- matched beam ($\sigma_0=120^\circ, \sigma=35^\circ$).}
\label{fig15}
\end{figure}

\begin{figure}[H]
\centering\epsfig{file=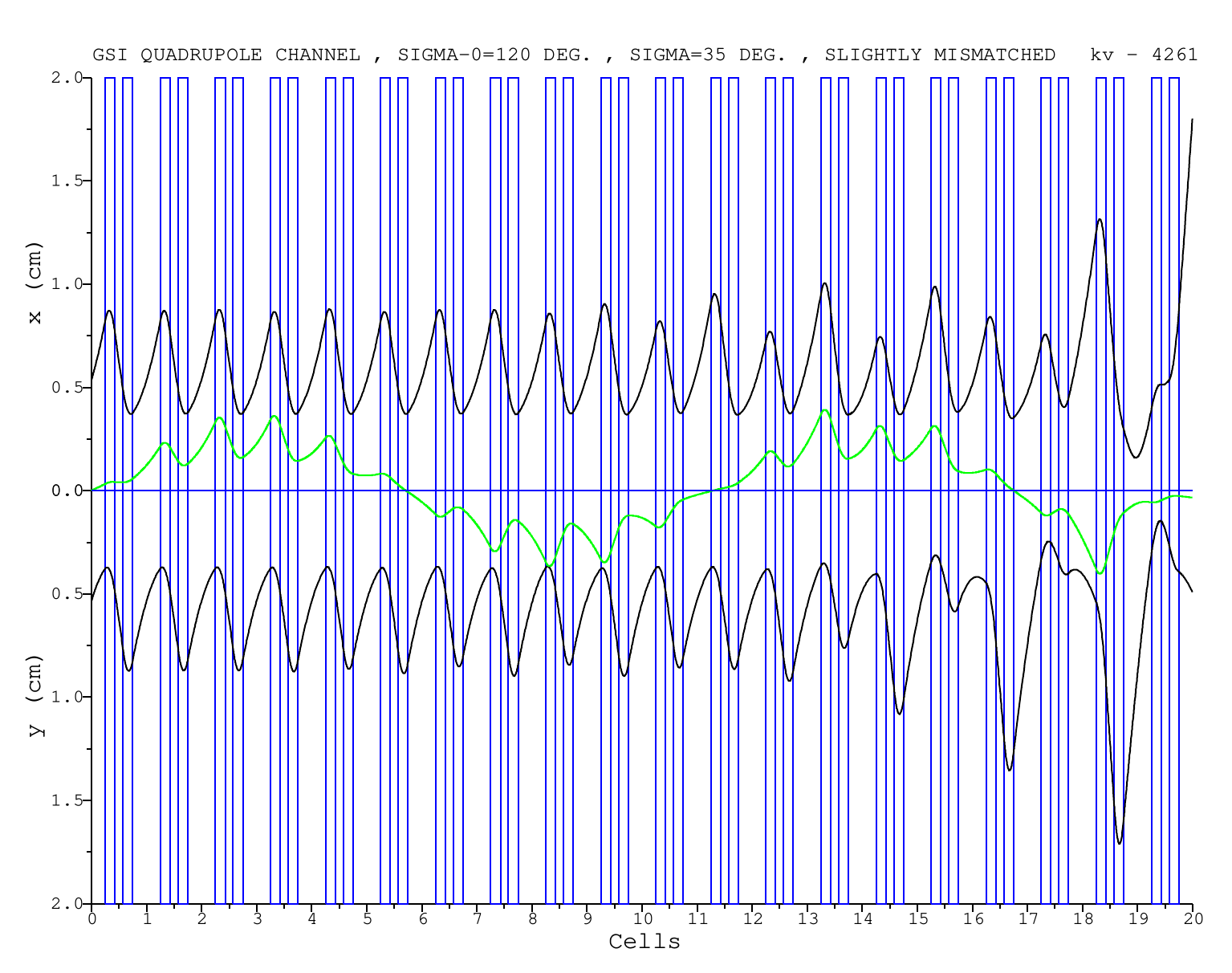,width=0.66\linewidth,angle=-90}
\caption{Quadrupole channel --- slightly mismatched beam ($\sigma_0=120^\circ, \sigma=35^\circ$).}
\label{fig16}
\end{figure}

In specific parameter intervals with $\sigma_0>90^\circ$, the expected envelope instability occurs.
An exactly matched beam still has periodic envelopes at unstable parameters, as illustrated by Fig.~\ref{fig15}.
Periodicity of this particular solution does not imply stability against perturbations.
The intended parameter values in the manuscript are $\sigma_0=120^\circ$ and $\sigma=35.1^\circ$, but the conflicting tune label in Fig.~\ref{fig15} requires confirmation.
In practice, finite matching errors can excite growing components, and rms matching alone does not ensure a periodic non-K-V distribution.
A slight mismatch with a component in the growing subspace initially leads to exponential growth of the envelope perturbation, as demonstrated in Fig.~\ref{fig16}.
For $\sigma_0=120^\circ$ and $\sigma=35.1^\circ$, the linearized calculation lies in a ``confluent'' resonance band (see Fig.~\ref{fig4})
\begin{displaymath}
\phi_1=198^\circ,\quad\phi_2=162^\circ,\quad|\lambda|=1.395.
\end{displaymath}
Thus the one-period matrix has a complex quartet with growing and contracting components, and a generic perturbation is unstable.

\section{PARTICLE SIMULATION RESULTS\label{sec5}}
\subsection{Solenoid Channel}
\begin{figure}[H]
\centering\epsfig{file=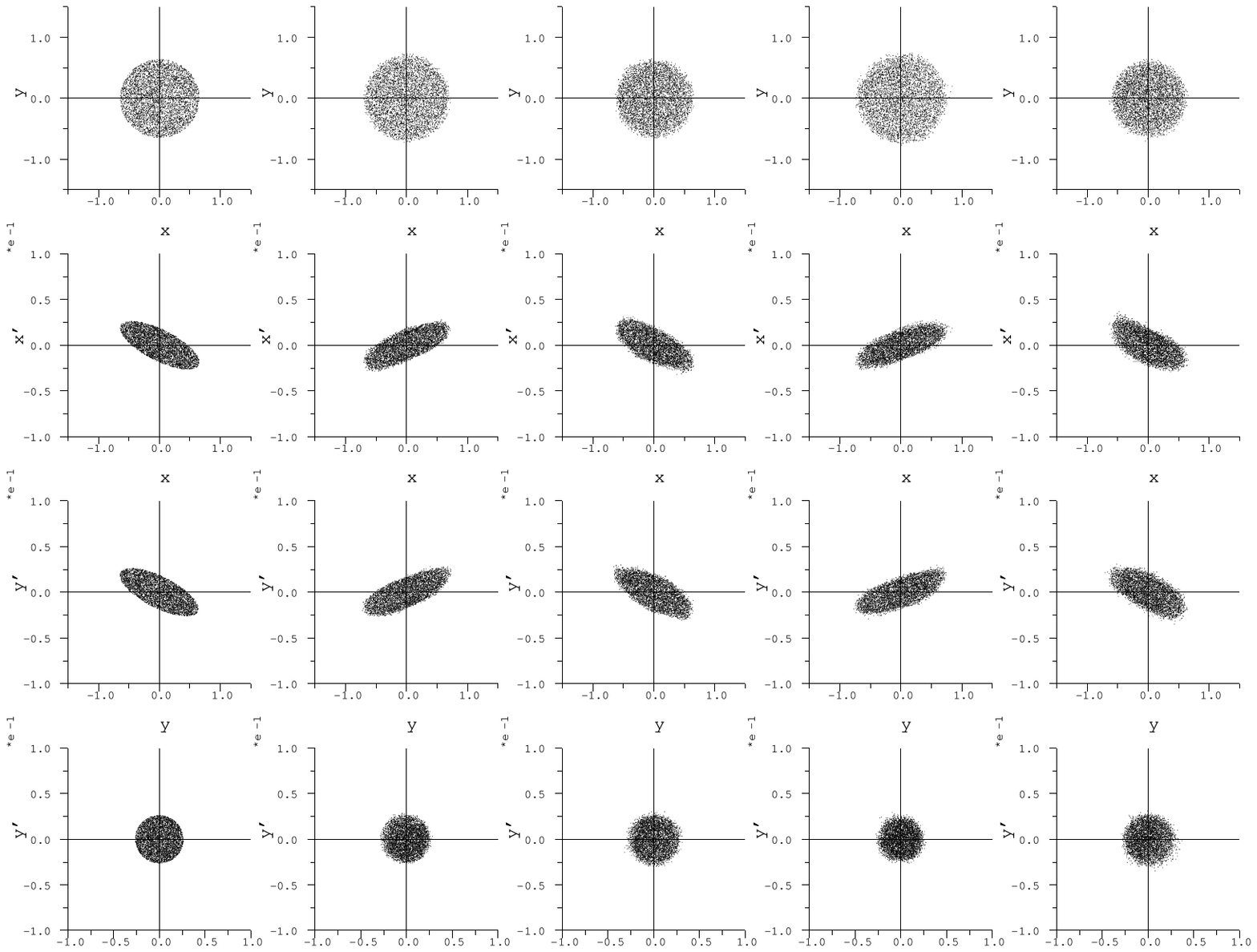,width=0.75\linewidth,angle=-90}
\\~\\ Period\hspace*{5mm} 0\hspace*{2cm} 10\hspace*{2cm} 20\hspace*{2cm} 30\hspace*{2cm} 40\hspace*{\fill}
\caption{Solenoid channel, particle simulation (K-V Dist.) --- ``in-phase'' mode ($\sigma_0=60^\circ$, \mbox{$\sigma=21^\circ$}).}
\label{fig17}
\end{figure}

\begin{figure}[H]
\centering\epsfig{file=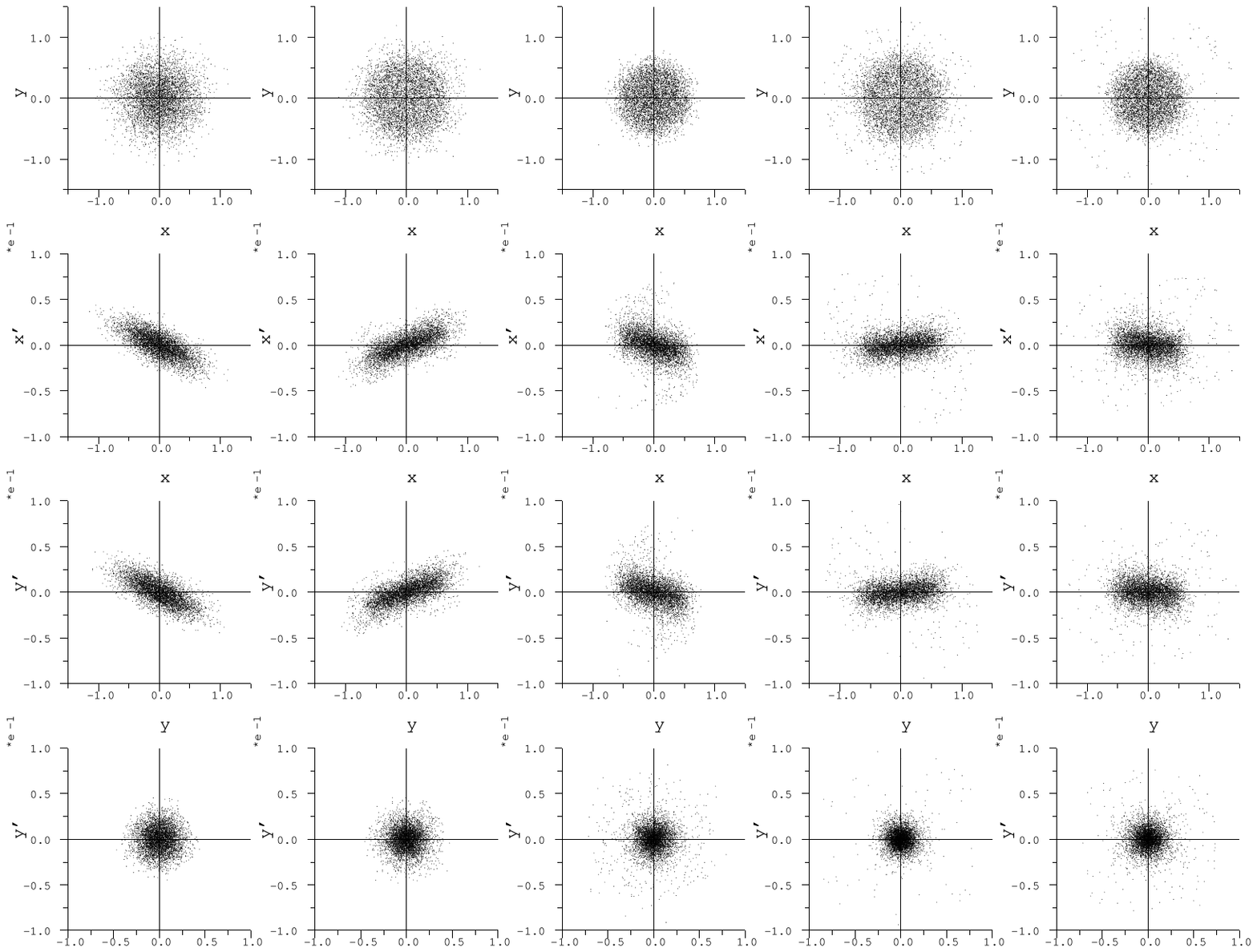,width=0.75\linewidth,angle=-90}
\\~\\ Period\hspace*{5mm} 0\hspace*{2cm} 10\hspace*{2cm} 20\hspace*{2cm} 30\hspace*{2cm} 40\hspace*{\fill}
\caption{Solenoid channel, particle simulation (Gauss-Dist.) --- ``in-phase'' mode (\mbox{$\sigma_0=60^\circ$}, $\sigma=21^\circ$).}
\label{fig18}
\end{figure}

\begin{figure}[H]
\centering\epsfig{file=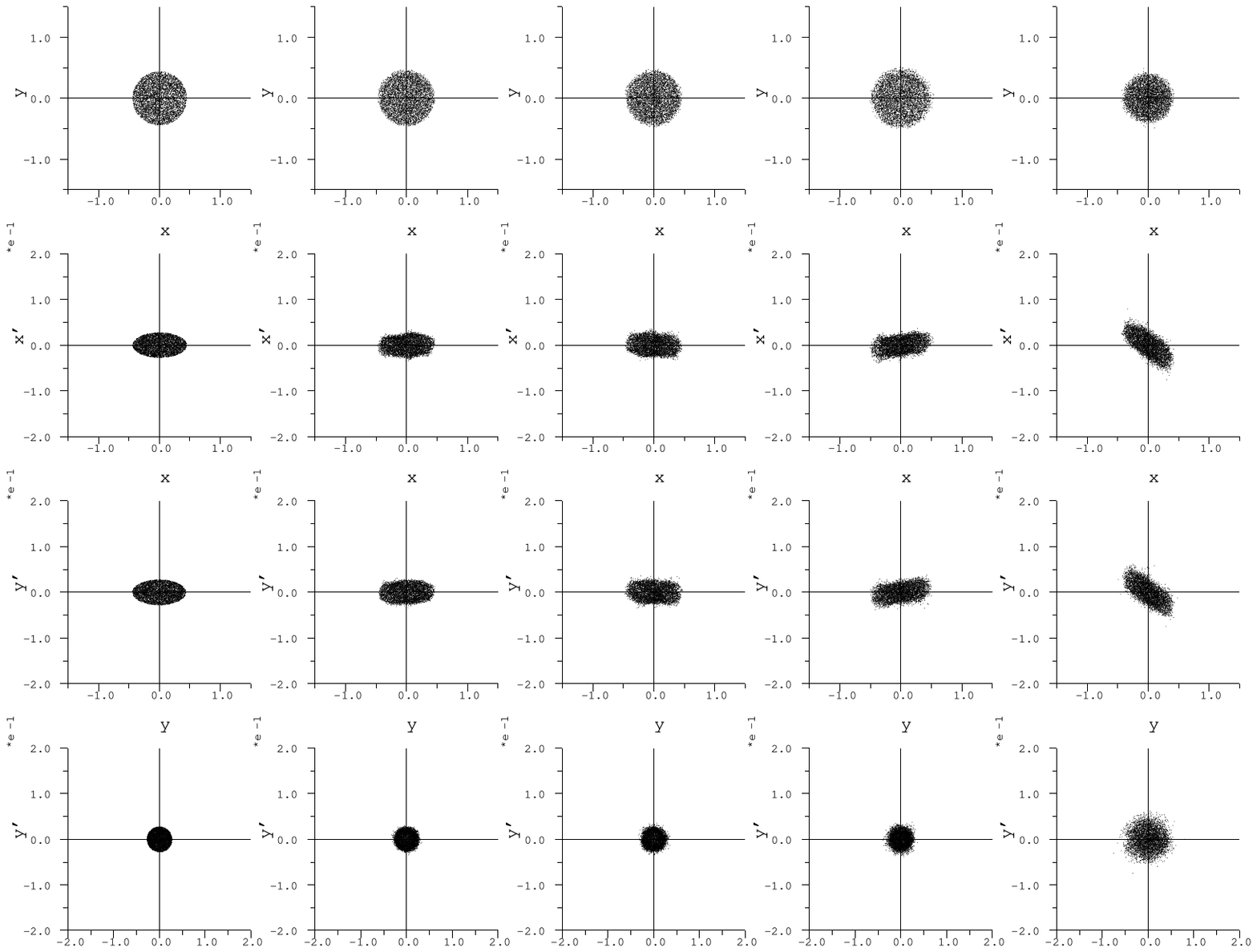,width=0.75\linewidth,angle=-90}
\\~\\ Period\hspace*{5mm} 0\hspace*{2cm} ~5\hspace*{2cm} 10\hspace*{2cm} 15\hspace*{2cm} 20\hspace*{\fill}
\caption{Solenoid channel, particle simulation (K-V Dist.) --- initially matched (\mbox{$\sigma_0=120^\circ$}, $\sigma=35^\circ$).}
\label{fig19}
\end{figure}

\begin{figure}[H]
\centering\epsfig{file=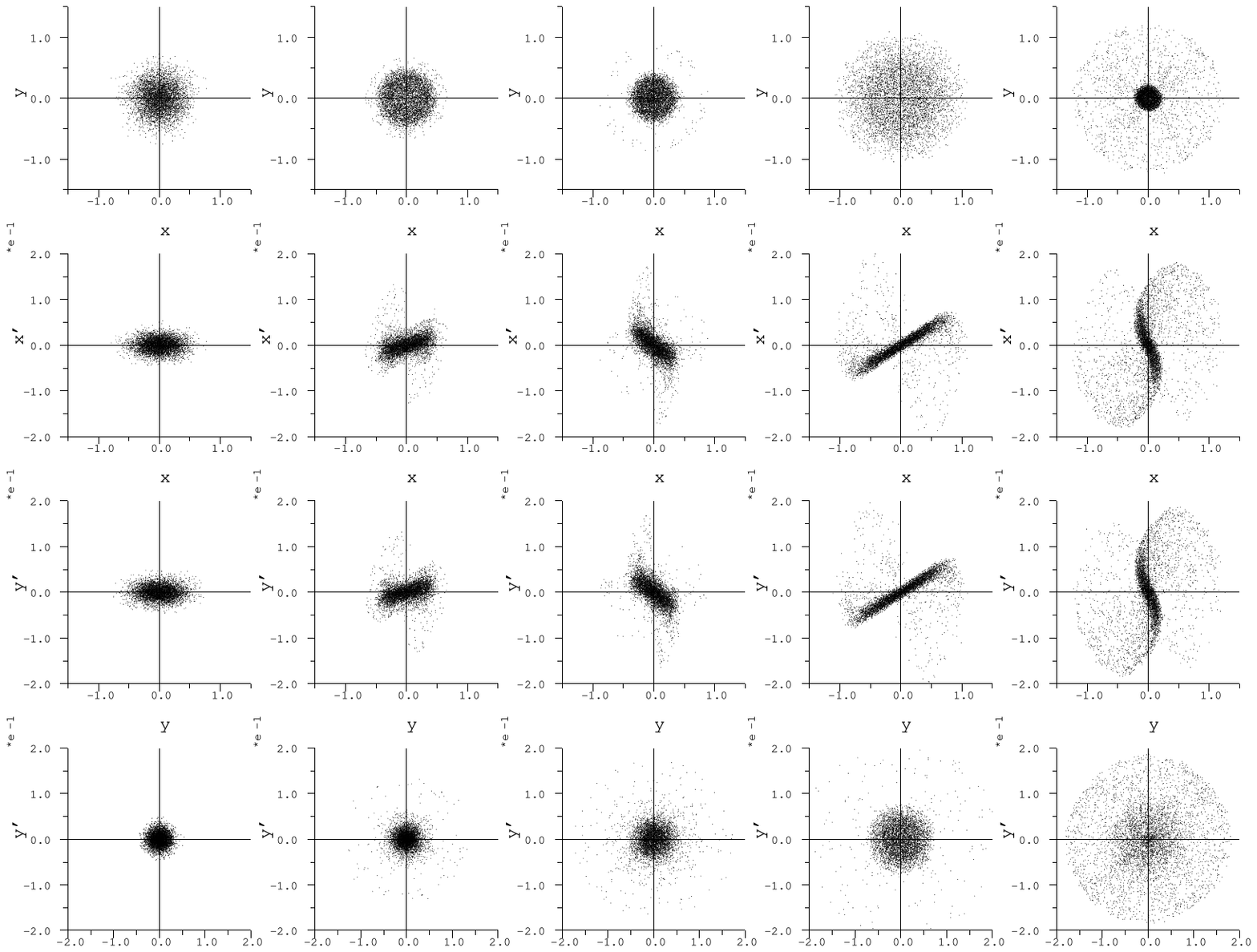,width=0.75\linewidth,angle=-90}
\\~\\ Period\hspace*{5mm} 0\hspace*{2cm} ~5\hspace*{2cm} 10\hspace*{2cm} 15\hspace*{2cm} 20\hspace*{\fill}
\caption{Solenoid channel, particle simulation (Gauss Dist.) --- initially rms-matched (\mbox{$\sigma_0=120^\circ$}, \mbox{$\sigma=35^\circ$}).}
\label{fig20}
\end{figure}

To obtain further insight into the beam physics of stable and unstable envelope oscillations, we performed particle-simulation studies with the modified PARMILA code at GSI, which features a fast $x$-$y$ Poisson solver to calculate the space-charge forces for long beams.
The beam was represented by $5000$ macroparticles, and in addition to a K-V distribution, we also studied the behavior of a more realistic Gaussian distribution in transverse phase space.
The figures display four phase-space projections of the particle distribution at fixed positions within the focusing cells.
Figures~\ref{fig17}, \ref{fig18}, \ref{fig21}, and~\ref{fig22} show periods $0,10,20,30,40$, whereas Figs.~\ref{fig19}, \ref{fig20}, \ref{fig23}, and~\ref{fig24} show periods $0,5,10,15,20$.
In the Maryland solenoid case, these projections are shown at the center of the drift space between the lenses, whereas for the GSI channel they are shown at the center of the larger of the two drift spaces between the quadrupole magnets.
For the Gaussian examples, the quoted $\sigma$ is used here to label the rms-equivalent matched K-V reference, not a common tune of all particles in the evolving nonlinear distribution.
An ``initially matched'' particle realization is matched to the accuracy of its initialization, rather than being an exact continuum periodic solution.
In the uncoupled K-V model, the single-particle forces are linear and the transfer map in each transverse phase plane has unit determinant.
It therefore preserves the rms emittance even when the envelope perturbations are unstable.
Emittance growth in the particle simulations describes dynamics beyond the fixed-emittance envelope equations.

Figure~\ref{fig17} shows the particle simulation of the ``in-phase'' mismatch case ($\sigma_0=60^\circ, \sigma=21.2^\circ$) for the Maryland solenoid channel using an initial K-V distribution.
The in-phase envelope oscillation has an approximate wavelength of four structure periods, as in Fig.~\ref{fig7}.
The ten-period spacing of the snapshots therefore corresponds to about two and a half envelope oscillations, so successive plots sample approximately opposite mismatch phases rather than the same phase.
This is consistent with the alternating tilts of the phase-space ellipses.
No appreciable rms emittance growth is observed in this simulation.

Figure~\ref{fig18} shows the same case using an initial Gaussian distribution.
Due to the nonlinear space-charge forces, we see aberration effects (here mostly in the $(y,y^\prime)$-phase-space projection), leading to a small transverse emittance growth of about $15\%$.
The initial Gaussian distribution can also relax because its density profile is not generally a self-consistent periodic equilibrium.
The reported growth therefore need not be due to envelope mismatch alone.
Separating these effects would require comparison with an otherwise identical, initially rms-matched Gaussian control.

For the case $\sigma_0=120^\circ, \sigma=34.6^\circ$, the particle simulation of an initially matched beam with a K-V distribution is shown in Fig.~\ref{fig19}.
The growth of the mismatch is consistent with the unstable envelope mode illustrated in Fig.~\ref{fig9}.
An exactly matched continuum K-V solution would remain periodic, but sampling fluctuations, initialization errors, and numerical field errors can seed the unstable mode in a finite-particle simulation.
The present plots do not identify the dominant seed.
The increasing mismatch is indicated by the fact that both transverse phase-space ellipses become more and more tilted.

The simulation results for an initial Gaussian distribution with identical initial rms parameters are plotted in Fig.~\ref{fig20}.
As is obvious from this figure, the behavior of a Gaussian beam is drastically different from the K-V case, due to the nonlinear space-charge forces.
Thus the Gaussian beam not only shows growing mismatch, but also a significant emittance growth by a factor greater than $2.6$ after the first $20$ cells.
This behavior is possibly also affected by the presence of higher-order instabilities.\cite{hofmann}

\subsection{Quadrupole Channel}
\begin{figure}[H]
\centering\epsfig{file=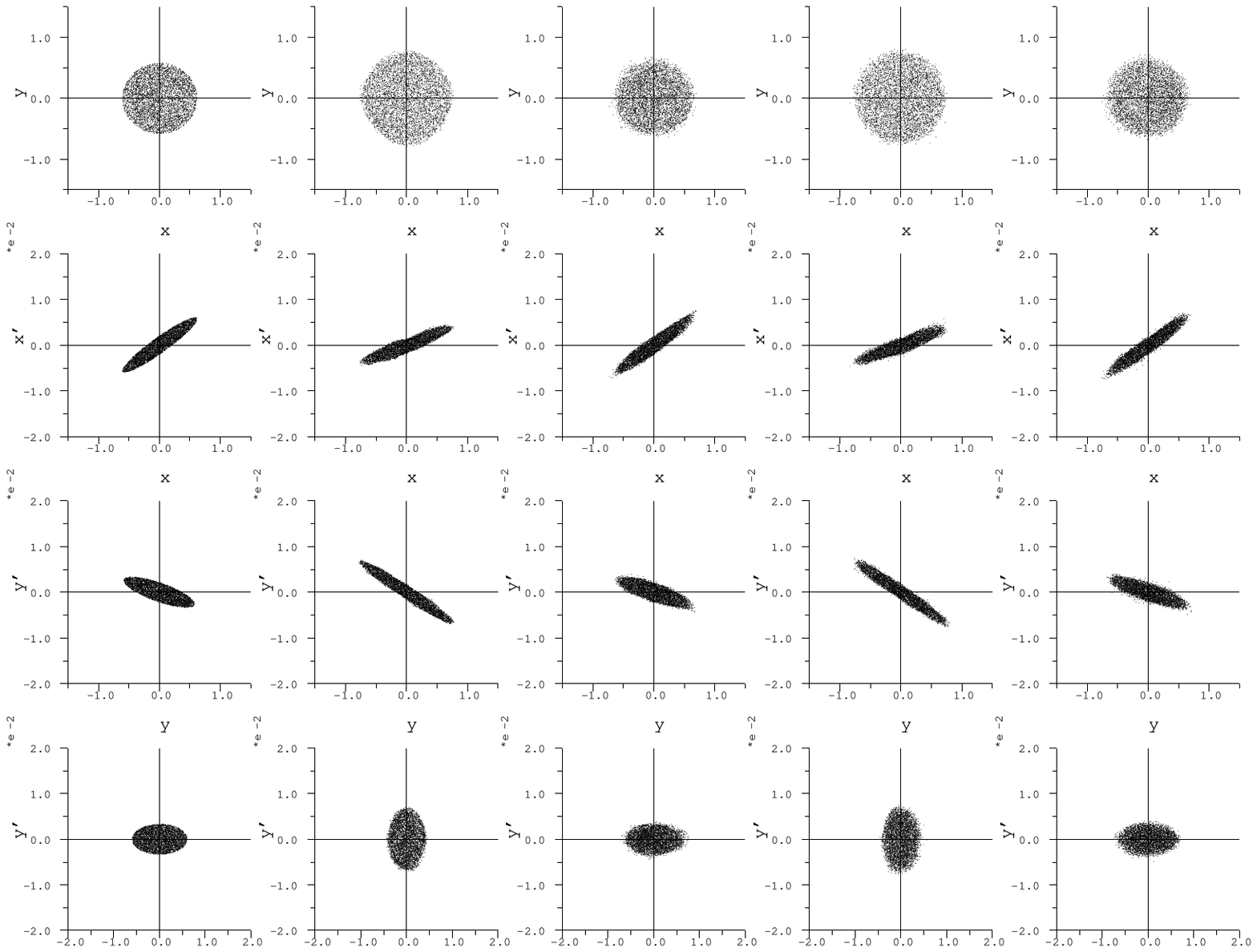,width=0.75\linewidth,angle=-90}
\\~\\ Period\hspace*{5mm} 0\hspace*{2cm} 10\hspace*{2cm} 20\hspace*{2cm} 30\hspace*{2cm} 40\hspace*{\fill}
\caption{Quadrupole channel, particle simulation (K-V Dist.) --- ``in-phase'' mode (\mbox{$\sigma_0=60^\circ$}, $\sigma=21^\circ$).}
\label{fig21}
\end{figure}

Figure~\ref{fig21} shows the particle simulation results of the ``in-phase'' mismatch case ($\sigma_0=60^\circ, \sigma=21^\circ$) for the GSI quadrupole channel using an initial K-V distribution.
It can be compared with the appropriate envelope plot of Fig.~\ref{fig13}.
No appreciable rms emittance growth is observed in this simulation.

\begin{figure}[H]
\centering\epsfig{file=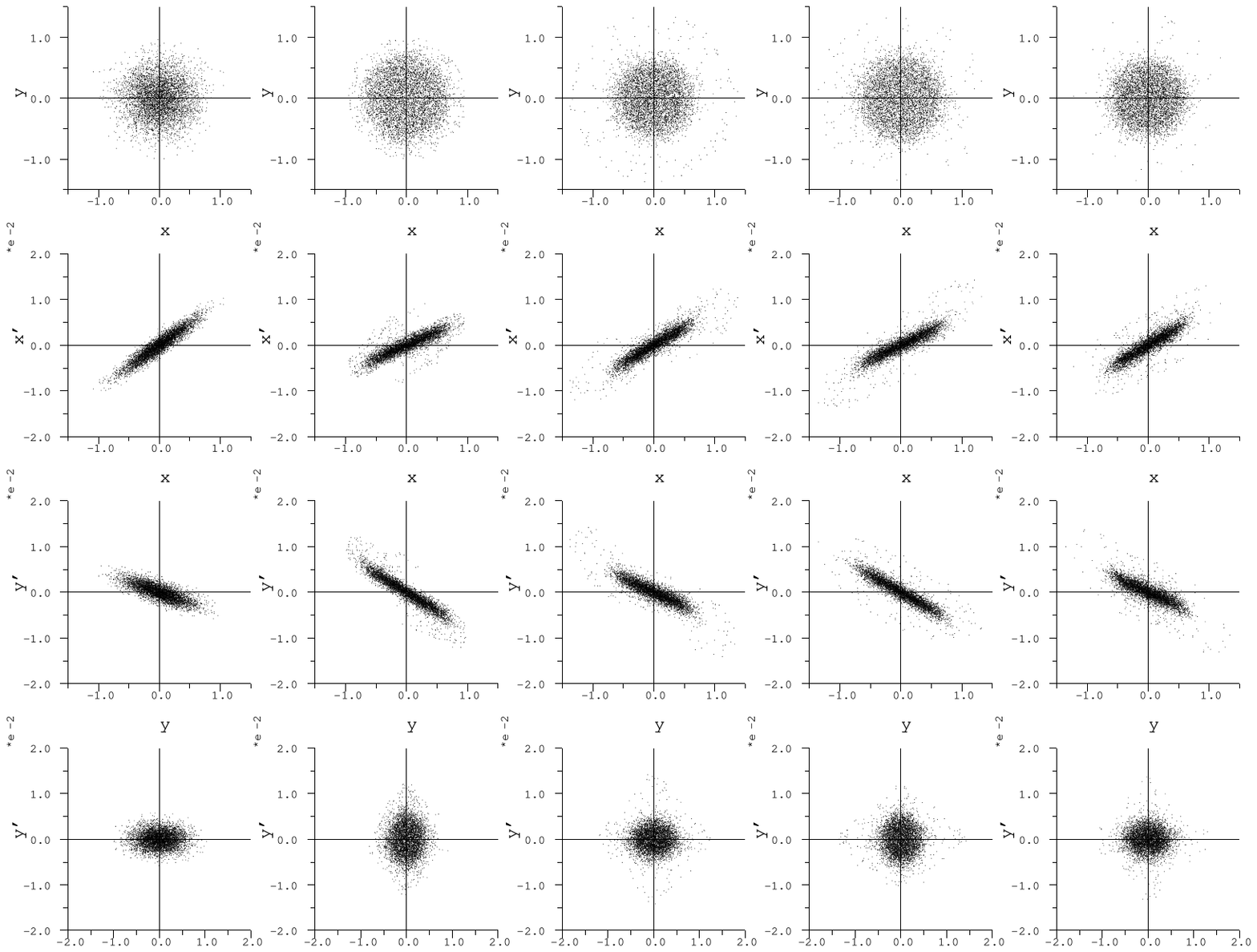,width=0.75\linewidth,angle=-90}
\\~\\ Period\hspace*{5mm} 0\hspace*{2cm} 10\hspace*{2cm} 20\hspace*{2cm} 30\hspace*{2cm} 40\hspace*{\fill}
\caption{Quadrupole channel, particle simulation (Gauss-Dist.) --- ``in-phase'' mode (\mbox{$\sigma_0=60^\circ$}, $\sigma=21^\circ$).}
\label{fig22}
\end{figure}

If we use an equivalent Gaussian transverse phase-space density distribution, we get an rms emittance growth of about $50\%$ after $20$ periods for the special case assumed here.
The growth is consistent with the nonlinear space-charge forces resulting from the nonuniform particle distribution, but the simulation does not separately quantify density-profile relaxation and envelope-mismatch effects.
Figure~\ref{fig22} shows the typical spherical-aberration patterns (the initially elliptical distributions in the $(x,x^\prime)$- and $(y,y^\prime)$-phase-space projections become $S$-shaped), which form a ``halo'' around the beam core after a few periods.

\begin{figure}[H]
\centering\epsfig{file=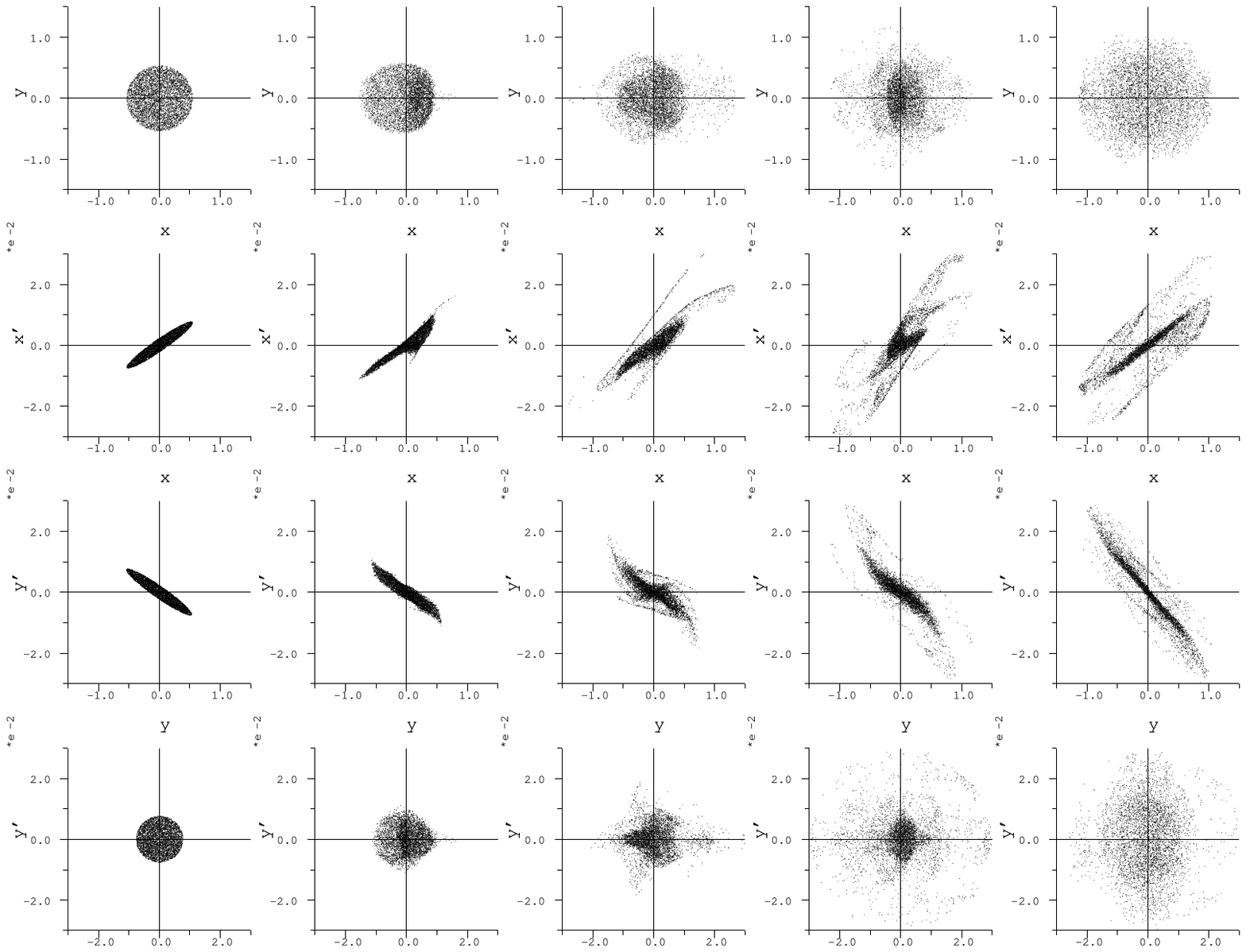,width=0.75\linewidth,angle=-90}
\\~\\ Period\hspace*{5mm} 0\hspace*{2cm} ~5\hspace*{2cm} 10\hspace*{2cm} 15\hspace*{2cm} 20\hspace*{\fill}
\caption{Quadrupole channel, particle simulation (K-V Dist.) --- initially matched (\mbox{$\sigma_0=120^\circ$}, $\sigma=35^\circ$).}
\label{fig23}
\end{figure}

\begin{figure}[H]
\centering\epsfig{file=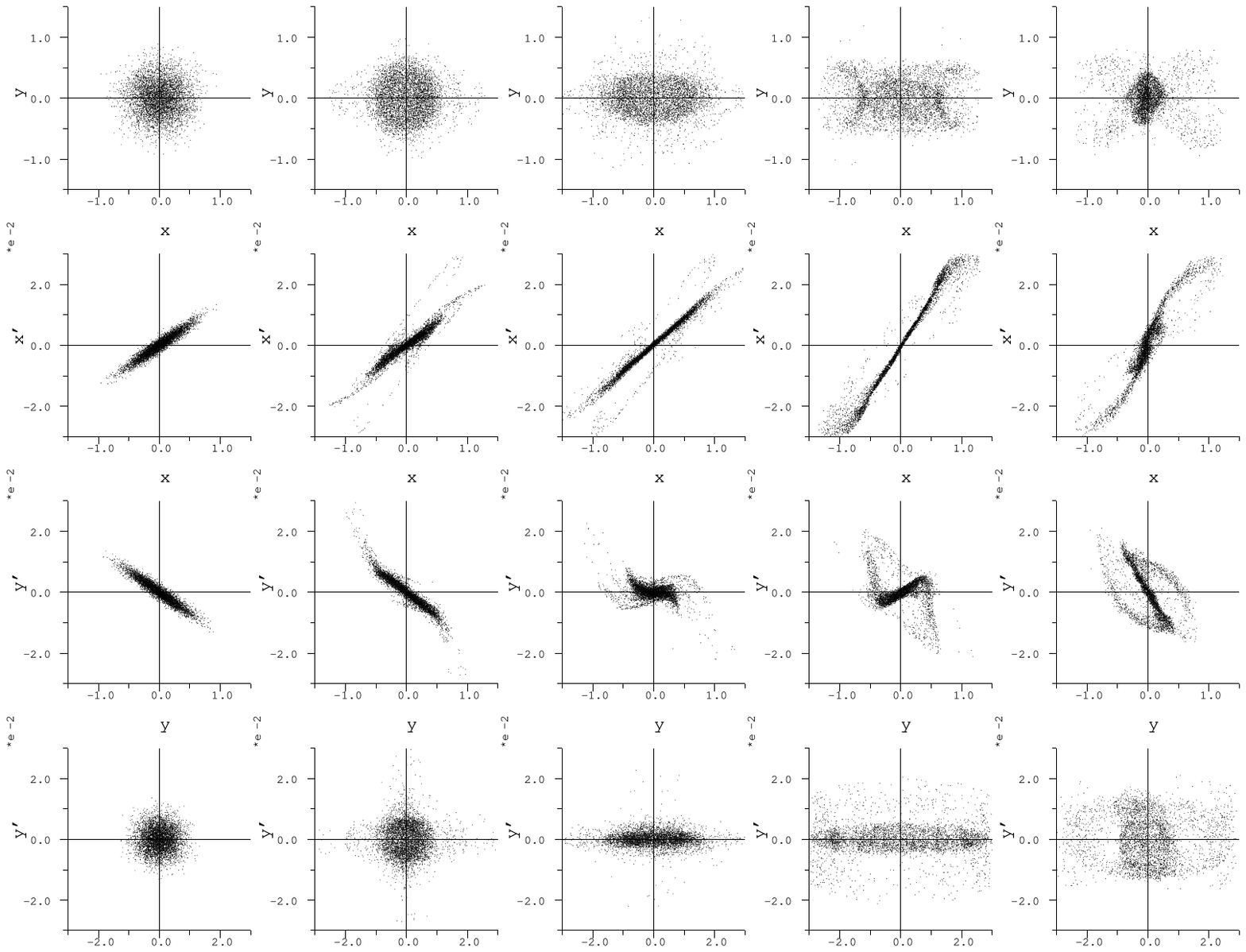,width=0.75\linewidth,angle=-90}
\\~\\ Period\hspace*{5mm} 0\hspace*{2cm} ~5\hspace*{2cm} 10\hspace*{2cm} 15\hspace*{2cm} 20\hspace*{\fill}
\caption{Quadrupole channel, particle simulation (Gauss Dist.) --- initially rms-matched (\mbox{$\sigma_0=120^\circ$}, $\sigma=35^\circ$).}
\label{fig24}
\end{figure}

In the region $\sigma_0>90^\circ$, we have simulated the case $\sigma_0=120^\circ$, $\sigma=35^\circ$ for both initial density distributions.
Figure~\ref{fig23} shows the initial K-V distribution, which gets seriously distorted after about $10$ periods, yielding an rms emittance growth factor of more than $3.5$ after $20$ periods.
This growth from an initially K-V distribution cannot be explained by the fixed-emittance envelope instability alone.
It indicates a departure from the assumed uncoupled K-V dynamics, which can involve distribution perturbations, nonlinear self-fields, other collective modes, or numerical effects.
Their relative contributions are not isolated by the phase-space plots.
The beam behavior becomes even worse for the Gaussian distribution illustrated in Fig.~\ref{fig24}.

The emittance-growth values reported here refer to the simulated cases.
No convergence study in macroparticle number, Poisson-grid resolution, or longitudinal integration step is documented here.
A quantitative assessment of numerical uncertainty therefore requires additional initialization and field-solver information and convergence tests.

\section{CONCLUSIONS}
A matched K-V beam in a periodic focusing channel is characterized by envelope radii and rms widths that repeat with the periodicity of the external focusing force.
Their period-averaged values are constant, but their rms widths need not be constant along the channel.
In the case of mismatch, the mean radius performs oscillations which according to the smooth-approximation theory presented in Section~\ref{sec2} can be described in terms of two fundamental modes, one in which the amplitudes of the oscillations in the two perpendicular transverse directions are ``$180^\circ$ out-of-phase'' and one in which they are ``in-phase''.
The phase advances of these two oscillation modes in one focusing period are found to be
\begin{displaymath}
\phi_1=\sqrt{\sigma_0{}^2+3\sigma^2}\quad\mathrm{and}\quad\phi_2=\sqrt{2\sigma_0{}^2+2\sigma^2},
\end{displaymath}
respectively.
A more accurate linear-perturbation analysis of the K-V envelope equations, in which the periodic variation is not smoothed out, reveals envelope-instability bands at specific values of $\sigma$ when $\sigma_0>90^\circ$.
The unstable bands shown are associated with $\phi_1=180^\circ$ or $\phi_2=180^\circ$ (``parametric resonance'' between envelope oscillations and the periodic focusing structure), or with $\frac{1}{2}(\phi_1+\phi_2)=180^\circ$ (``confluent resonance'').
Exponential instability is established by a Floquet multiplier outside the unit circle, not by these phase relations alone, and depends on both $\sigma_0$ and $\sigma$ (i.e., beam intensity).
For the examples studied, the maximum growth factors increase with $\sigma_0$ and are larger in the quadrupole channel than in the solenoid channel.
These results are in agreement with the studies of Ref.~4, where an alternative analysis based on the Vlasov equation was used.
In the region below $\sigma_0=90^\circ$, which is stable with regard to the envelope modes analyzed here, the smooth-approximation results for their frequencies (or phase angles) agree closely with the periodic-coefficient calculation.
This does not exclude other collective modes or emittance growth from non-K-V dynamics.
Numerical integration of the full nonlinear K-V equations supports the first-order predictions for small envelope perturbations.
The particle simulations illustrate the corresponding coherent behavior, but also show distribution distortion and emittance growth that the fixed-emittance envelope theory does not predict.
Differences between K-V and Gaussian distributions can be attributed for the most part to the nonlinear space-charge forces in the latter case, which show typical $S$-shaped arms of spherical aberrations in the phase-space plots.
Formulas for calculating $\sigma_0$ and $\sigma$ are given in Ref.~3.
For convenience we repeat the ``smooth-approximation'' result for $\sigma$
\begin{displaymath}
\sigma=\sigma_0\cdot(\sqrt{1+u^2}-u),
\end{displaymath}
where
\begin{displaymath}
u=\frac{KS}{2\epsilon\sigma_0}.
\end{displaymath}
Here $\sigma_0$ in the definition of $u$ is in radians.
We found in our studies that this formula agrees remarkably well with the numerical periodic-envelope results in the region of practical interest for beam transport ($\sigma_0\leq90^\circ$).
\vspace*{-5mm}
\acknowledge
This work was performed while one of us (M.~R.) was on sabbatical leave from the University of Maryland at GSI, W.~Germany.
The support from the Alexander von Humboldt Foundation and the hospitality of the GSI laboratory during this leave are gratefully acknowledged.
We would also like to thank J.~Klabunde for valuable discussions.

\end{document}